\documentclass[10pt,twocolumn,oneside,final]{IEEEtran}

\usepackage{cite}
\usepackage{graphicx}
\usepackage{amsmath}
\usepackage{times}
\usepackage{latexsym}
\usepackage{bm}
\usepackage{amssymb}
\usepackage[center]{caption2}
\usepackage{array}
\usepackage{setspace}
\usepackage{fancyhdr}
\usepackage{cite,graphicx,amsmath,amssymb}
\usepackage{citesort}
\usepackage{color}
\usepackage{multirow}
\usepackage{times}
\usepackage{amsmath}
\usepackage{amssymb}
\usepackage{amsfonts}
\usepackage{graphics}
\usepackage{amssymb}
\usepackage{booktabs}

\ifCLASSINFOpdf
\else
\fi

\makeatother

\begin{document}

\title{Power Allocation for Multi-Pair Massive MIMO Two-Way AF Relaying with Linear Processing}
\author{Yongyu Dai,~\IEEEmembership{Student Member,~IEEE,}
and~Xiaodai Dong,~\IEEEmembership{Senior~Member,~IEEE,}
\thanks{
Y.~Dai and X.~Dong are with the Department of Electrical and Computer Engineering, University of Victoria, Victoria, BC, Canada (email: yongyu@uvic.ca,~xdong@ece.uvic.ca).
}}
\maketitle

%

\begin{abstract}
   In this paper, we consider a multi-pair two-way amplify-and-forward relaying system where multiple sources exchange information via a relay node equipped with large-scale antenna arrays. Given that channel estimation is non-ideal, and that the relay employs either maximum-ratio combining/maximum-ratio transmission (MRC/MRT) or zero-forcing reception/zero-forcing transmission (ZFR/ZFT) beamforming, we derive two corresponding closed-form lower bound expressions for the ergodic achievable rate of each pair sources. The closed-form expressions enable us to design an optimal power allocation (OPA) scheme that maximizes the sum spectral efficiency under certain practical constraints. As the antenna array size tends to infinity and the signal to noise ratios become very large, asymptotically optimal power allocation schemes in simple closed-form are derived. The capacity lower bounds are verified to be accurate predictors of the system performance by simulations, and the proposed OPA outperforms equal power allocation (EPA). It is also found that in the asymptotic regime, when MRC/MRT is used at the relay and the link end-to-end large-scale fading factors among all pairs are equal, the optimal power allocated to a user is inverse to the large-scale fading factor of the channel from the user to the relay, while OPA approaches EPA when ZFR/ZFT is adopted.
\end{abstract}

\IEEEpeerreviewmaketitle

\section{Introduction}\label{sec:introduction}

Massive multiple-input multiple-output (MIMO) transmission, in which a base station is equipped with hundreds of antennas for multiuser operation, is considered as one of the key enabling technologies for 5G~\cite{5G}. In \cite{Marzetta2010}, it was first proposed for multi-cell noncooperative scenarios. Such large antenna arrays can substantially reduce the effects of noise, small-scale fading and inter-user interference, using only simple signal processing techniques with reduced total transmit power, and only inter-cell interference caused by pilot contamination remains~\cite{Marzetta2010,FF2013}. Subsequently, the energy and spectral efficiency of very large multiuser MIMO systems were investigated in the single cell scenarios in \cite{Ngo2013}, which showed that the power radiated by the terminals could be made inversely proportional to the square-root of the number of base station antennas with no reduction in performance when considering imperfect channel state information (CSI), and that the power could be made inversely proportional to the number of antennas if perfect CSI were available.

Currently, massive MIMO combined with cooperative relaying is considered as a strong candidate for the development of future energy-efficient networks and has received increasing attention~\cite{Ngo22013,Suraweera2013,Cui2014,Liu2014,XSL2014,Ngo2014}. In the field of cooperative relaying, two-way relaying technique outperforms one-way relaying in terms of spectral efficiency, since it employs the principle of network coding at the relay in order to mix the signals received simultaneously from two links for subsequent forwarding, and then applies the self-interference cancellation (SIC) at each user to extract the desired information~\cite{Nosratinia2004,Rankov2007}. For the multi-pair two-way relaying with massive MIMO, \cite{Cui2014} obtained the asymptotic spectral and energy efficiencies of the system analytically with both maximum-ratio combining/maximum-ratio transmission (MRC/MRT) and zero-forcing reception/zero-forcing transmission (ZFR/ZFT) beamforming, supposing that the number of relay antennas approaches to infinity and the transmit power of all users is equal. However, only asymptotic cases with perfect CSI and perfect SIC were studied and no closed-form expression for the ergodic achievable rate with finite number of relay antennas was derived in \cite{Cui2014}. In \cite{XSL2014}, the ergodic achievable rates were investigated with perfect CSI based MRC/MRT used at the relay, providing a capacity lower bound, the derivation of which involved asymptotic approximations. Neither \cite{Cui2014} nor \cite{XSL2014} considers imperfect CSI or power allocation (PA) problems.

In the literature, instantaneous power allocation schemes based on instantaneous rate for regular scale MIMO rather than massive MIMO were presented for one way or two way AF wireless relay systems to improve system performance~\cite{KH2009,D2011,Cao2013}. In massive MIMO systems, ergodic rate is usually used in power allocation because the instantaneous rate approaches the ergodic rate as the number of antennas tends to infinity due to the law of large numbers, and such PA schemes are more practical with lower complexity than instantaneous rate based ones. In \cite{Ngo2014}, an ergodic rate based optimal power allocation (OPA) scheme was proposed for a multi-pair decode-and-forward (DF) one-way relaying with massive arrays. Nevertheless, power allocation has not been addressed in a massive MIMO two-way relaying system. Besides, there is no closed-form ergodic rate expressions derived for massive MIMO two-way relaying with ZFR/ZFT in the literature.

This paper considers a multi-pair two-way AF relaying system where multiple sources exchange information via a relay node equipped with large-scale arrays. Assuming imperfect CSI estimation, the relay station employs the MRC/MRT and ZFR/ZFT beamforming to process the signals, respectively. First, utilizing the technique in \cite{Marzetta2006,Jose2011}, we derive for the first time two statistical CSI (SCSI) based closed-form lower bounds for the ergodic achievable rate in the case of arbitrary number of relay antennas (without resorting to asymptotic approximations) with MRC/MRT and ZFR/ZFT processing, respectively, based on the properties of Wishart and inverse Wishart matrices. Having obtained the closed-form expressions, we are able to design an OPA scheme that maximizes the sum spectral efficiency under certain practical constraints. The proposed OPA scheme is based on geometric programming (GP)~\cite{convex2004,KH2009}, which can be solved by conventional optimization tools, such as CVX~\cite{cvx2008}. Considering the massive MIMO properties, an asymptotically OPA is presented for the asymptotic regimes with closed-form solutions. The derived closed-form expressions for the achievable rate are verified to be accurate predictors of the system performance by Monte-Carlo simulations. Furthermore, in order to demonstrate the effectiveness of the developed OPA schemes, simulations of spectral efficiency are conducted under different system configurations, respectively, in comparison to the equal power allocation (EPA) schemes.

The rest of the paper is organized as follows. We briefly describe the system model for the multi-pair two-way AF relaying in Section~\ref{sec:sys_model}. In Section~\ref{Achiev_Rate}, two closed-form expressions for the achievable rate are derived for MRC/MRT and ZFR/ZFT, respectively, followed by asymptotic analysis. Then, an OPA and an asymptotically OPA are proposed by solving the sum-rate maximization based optimization problem in Section~\ref{PA}. Furthermore, simulation results under different system configurations are given in Section~\ref{NumericalResults} to demonstrate the effectiveness of both derived rate expressions and developed OPAs. Finally, we draw our conclusions in Section~\ref{conclusion}.

\emph{Notations:} For a matrix $\mathbf{X}$, we use ${\rm Tr}\{\bf{X}\}$, ${\bf{X}}^T$, ${\bf{X}}^H$ and ${\bf{X}}^*$ to denote the trace, the transpose, the Hermitian transpose, and the conjugate, respectively. The symbol $\left\|{\bf x}\right\|$ indicates the 2-norm of vector $\bf x$ and $diag\{{\bf x}\}$ denotes a diagonal matrix with $\bf x$ being its diagonal entries. Moreover, ${\bf I}_N$ denotes the $N\times N$ identity matrix, ${\rm E}[\cdot]$ and ${\rm Var}[\cdot]$ denotes the expectation and the variance operators, respectively. $\left[a\right]^+$ denotes ${\rm max}\left\{a,0\right\}$. Finally, ${\bf{x}} \sim \mathcal{CN}(0,{\bf D}_x)$ represents a circularly symmetric complex Gaussian vector $\bf x$ with zero mean and covariance matrix ${\bf D}_x$.

\section{System Model}\label{sec:sys_model}
\begin{figure}
\centering
\includegraphics[scale=0.32]{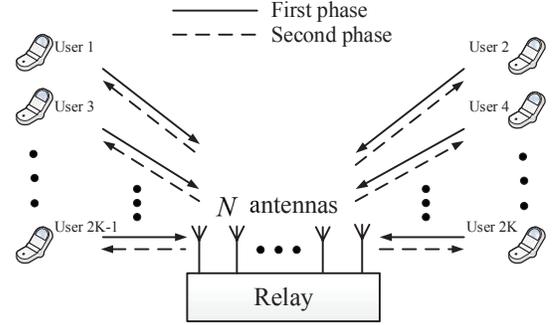}
\caption{System diagram of multi-pair two-way AF relaying.}
\label{fig:SystemModel}
\end{figure}

Fig.~\ref{fig:SystemModel} shows the considered multi-pair two-way AF relaying network, where $K$ pairs of users communicate with the help of a common relay station by sharing the same time-frequency resources. In this system, two single-antenna users in the $l$th user pair denoted by $(2l-1,2l)$ or $(2l,2l-1)$ (for $l=1,\cdots,K$) want to exchange information with each other via the relay equipped with $N$~($N\gg 2K\gg 1$) antennas. Notably, the direct links between the corresponding users are assumed non-existing in the two-way relaying system. Typically, a two-way network is divided into two phases, namely the multiple-access (MA) phase and the broadcast (BC) phase \cite{Nosratinia2004}. In the MA phase, information is sent from the user pairs to the relay; while in the BC phase, the relay broadcasts the processed information.

Let $p_i$~($i = 1,2,\cdots,2K$) and ${P_{\rm R}}$ denote the power transmitted by user $i$ and the relay corresponding to the MA and BC phases, respectively. We assume that all the channels between the users and the relay follow independent and identically distributed (i.i.d.) Rayleigh fading and time division duplex (TDD) is adopted in all transceivers. Thus, supposing that ${\bf{g}}_i\in{\mathbb{C}^{N \times 1}}~(i= 1,2,\cdots,2K)$ is the channel between the $i$th user and the relay, ${\bf{g}}_i$ contains the i.i.d. $\mathcal{CN}(0,\sigma_i^2)$ elements, where $\sigma_i^2$ represents the corresponding large-scale fading coefficient. In this way, we can denote the channel matrix between all the users and the relay accounting for both small-scale fading and large-scale fading by
\begin{equation}\label{channel_matrix}
\begin{split}
&{\bf{G}} = {\bf{H}}{{\bf{D}}^{{1 \mathord{\left/
 {\vphantom {1 2}} \right.
 \kern-\nulldelimiterspace} 2}}} = \left[ {{{\bf{g}}_1},{{\bf{g}}_2}, \ldots ,{{\bf{g}}_{2K}}} \right] \in {\mathbb{C}^{N \times 2K}}
\end{split}
\end{equation}
where ${\bf{H}}\in \mathbb{C}^{N \times 2K}$ includes the i.i.d. $\mathcal{CN}(0,1)$ small-scale fading coefficients, and ${\bf D}$ is the large-scale fading diagonal matrix with the $i$th diagonal elements denoted by $\sigma_i^2$~($i=1,2,\cdots,2K$).

\subsection{Channel Estimation}
Practically, the channel matrices in both the MA and BC phases have to be estimated for relay processing. However, due to the large-scale antenna array at the relay, channel estimation at the user side becomes rather impractical. Thus, time division duplex (TDD) is adopted here and channel reciprocity can be utilized, i.e., only channel matrix ${\bf{G}}$ between all the users and the relay has to be estimated based on the uplink training. The relay then has the estimated CSIs of all uplink and downlink channels. The required channel related information at the user side can be calculated by the relay and fed back to the users, as will be explained later. At the beginning of each coherence interval $T$, all users simultaneously transmit pilot sequences of length $\tau $ symbols. The pilot sequences of all the $2K$ users are pairwisely orthogonal, i.e., $\tau  \ge 2K$ is required. Then the training matrix received at the relay is
\begin{equation}\label{train_BS}
\begin{split}
{{\bf{Y}}_{\rm{R}}} = \sqrt {\tau {p_{\rm P}}} {\bf{G\Phi }} + {{\bf{N}}_{\rm{R}}}
\end{split}
\end{equation}
where $p_{\rm P}$ is the transmit power of each pilot symbol, ${\bf{N}}_{\rm{R}}$ is the additive white Gaussian noise (AWGN) matrix with i.i.d. components following $\mathcal{CN}(0,\sigma_n^2)$, the training vector transmitted by the $i$th~($i=1,\cdots,2K$) user is denoted by the $i$th row of ${{\bf{\Phi }}}\in{\mathbb{C}^{2K\times \tau}}$, satisfying ${\bf{\Phi }}{\bf{\Phi }}^H={{\bf{I}}_{2K}}$. Moreover, since the rows of pilot sequence matrices are pairwisely orthogonal, we have ${{\bf{\phi }}^{(i)}}\left({{\bf{\phi }}^{(j)}}\right)^H=0$~($\forall i\ne j\in\{1,\dots,2K\}$).

In order to estimate the channel matrices ${\bf{G}}$, we employ the minimum mean-square-error (MMSE) estimation at the relay. The MMSE channel estimates are given by~\cite{Kay1993}
\begin{equation}\label{estimated_CSI}
\begin{split}
{\bf{\hat G}} = \frac{1}{{\sqrt {\tau {p_{\rm{P}}}} }}{{\bf{Y}}_{\rm{R}}}{{\bf{\Phi }}^H}{\bf{\tilde D}} = {\bf{G\tilde D}} + \frac{1}{{\sqrt {\tau {p_{\rm{P}}}} }}{{{\bf{\tilde N}}}_{\rm{R}}}{\bf{\tilde D}}
\end{split}
\end{equation}
where we define ${\bf{\tilde D}} \buildrel \Delta \over = {\left( {\frac{{{{\bf{D}}^{ - 1}}\sigma _n^2}}{{\tau {p_{\rm{P}}}}} + {{\bf{I}}_{2K}}} \right)^{ - 1}}$ and ${{{\bf{\tilde N}}}_{\rm{R}}} \buildrel \Delta \over = {{\bf{N}}_{\rm{R}}}{{\bf{\Phi }}^H}$. According to the property of ${\bf{\Phi }}$, we conclude that ${{{\bf{\tilde N}}}_{\rm{R}}}$ is composed of i.i.d. $\mathcal{CN}(0,\sigma_n^2)$ elements. Then,
\begin{equation}\label{estimated_CSI_exp}
{\bf{\hat G}} = {\bf{G}} + {\bf{\Xi }}=\left[ {{{\hat{\bf g}}_1},{{\hat{\bf g}}_2}, \ldots ,{{\hat{\bf g}}_{2K}}} \right] \in {\mathbb{C}^{N \times 2K}}
\end{equation}
where ${\bf{\Xi }}=\left[ {{{\bf{\xi }}_1},{{\bf{\xi }}_2}, \ldots ,{{\bf{\xi }}_{2K}}} \right]$ denotes the estimation error matrix which is independent of ${{\bf{\hat G}}}$ from the property of MMSE channel estimation~\cite{Kay1993}. Hence, we have ${{\bf{\hat G}}} \sim CN\left( 0,{{\bf{\hat D}}} \right)$ with ${{\bf{\hat D}}} = diag\left\{ {\hat \sigma _1^2,\hat \sigma _2^2, \ldots ,\hat \sigma _{2K}^2} \right\}$, and ${\bf{\Xi }} \sim CN\left( {0,{\bf{D}} - {\bf{\hat D}}} \right)$ with ${\bf{D}} - {\bf{\hat D}} = diag\left\{ {\sigma _{{\xi _1}}^2,\sigma _{{\xi _2}}^2, \ldots ,\sigma _{{\xi _{2K}}}^2} \right\}$. The diagonal elements satisfy $\hat \sigma _i^2 = \frac{{\tau {p_{\rm{P}}}\sigma _i^4}}{{\tau {p_{\rm{P}}}\sigma _i^2 + \sigma _n^2}}$ and $\sigma _{{\xi _i}}^2 \buildrel \Delta \over = \sigma _i^2 - \hat \sigma _i^2 = \frac{{\sigma _i^2}}{{\tau {p_{\rm{P}}}\sigma _i^2 + \sigma _n^2}}$ with $i=1,\dots,2K$.

\subsection{Data Transmission}
Since the relay station estimates all the channels, it employs linear processing MRC/MRT and ZFR/ZFT based on the imperfect CSI. While each user only has the knowledge of its pairwise effective channel coefficient for data detection and self-interference cancellation coefficient for SIC, which are calculated and sent out by the relay. In the MA phase, all the users transmit their signals simultaneously to the relay. That is, the received signal at the relay station is given by
\begin{equation}\label{transmit_signal}
\begin{split}
&{\bf{r}} = \sum\limits_{i = 1}^{2K} {\sqrt {{p_i}} {{\bf{g}}_i}{x_i}}  + {{\bf{n}}_{\rm{r}}} = {\tilde{\bf G}}{\bf x} + {{\bf{n}}_{\rm{r}}}
\end{split}
\end{equation}
where ${\tilde{\bf G}} = {\bf{GP}}$, ${\bf{P}} = diag\left\{ {\sqrt {{p_1}}, \sqrt {{p_2}}\ldots \sqrt {{p_{2K}}} } \right\}$ with each power satisfying $0\le p_i \le {P_0}^1$, ${\bf{x}} = {\left[ {{x_1},{x_2}, \ldots ,{x_{2K}}} \right]^T}$ with the $i$th element $x_i$ representing the transmitted signal by the $i$th user and ${\rm E}\left[ {{\bf{x}}{{\bf{x}}^H}} \right] = {{\bf{I}}_{2K}}$, ${\bf r}\in \mathbb{C}^{N\times 1}$, and ${{\bf{n}}_{\rm{r}}}$ is the additive white Gaussian noise (AWGN) vector at the relay with zero mean and the variance of $\sigma _n^2$.

Then, in the BC phase, the relay multiplies the received signal by a linear receiving and precoding matrix to yield the relay transmitted signal given by $\hat {\bf{r}} = {\bf{Fr}} \in {\mathbb{C}^{N \times 1}}$, where ${\bf F}$ is the combined beamforming matrix at the relay and its expression will be given in the next subsection. The transmitted signal satisfies the expected transmit power constraint at the relay~\cite{Hong2013}, i.e.,
\begin{equation}\label{power_consR}
\begin{split}
{P_{\rm{R}}} &= {\rm E}\left[ {{{\left\| \hat {\bf{r}} \right\|}^2}} \right] = {\rm Tr}\left\{ {{\rm E}\left[ {{\bf{F}}\left( {{\tilde{\bf G}}{{{\tilde{\bf G}}}^H} + \sigma _n^2{\bf{I}}_N} \right){{\bf{F}}^H}} \right]} \right\}.
\end{split}
\end{equation}
with a total power constraint $\sum\limits_{i = 1}^{2K} {{p_i}} + {P_{\rm{R}}}\le P$\footnote{Here, $P$ and $P_0$ are two constants preset for the total power constraint and individual power constraint, respectively.}. In the BC phase, the received signal at the $k'$th user can be expressed as
\begin{equation}\label{receive_signal}
\begin{split}
&{y_{k'}} = {\bf{g}}_{k'}^T\hat {\bf{r}} + {n_{k'}} = \underbrace{\sqrt {{p_k}} {\bf{g}}_{k'}^T{\bf{F}}{{\bf{g}}_k}{x_k}}_{\text{desired signal}} + \underbrace{\sqrt {{p_{k'}}} {\bf{g}}_{k'}^T{\bf{F}}{{\bf{g}}_{k'}}{x_{k'}}}_{\text{self-interference}} \\
&~~~+ \underbrace{{\bf{g}}_{k'}^T{\bf{F}}\sum\limits_{i \ne k,k'}^{2K} {\sqrt {{p_i}} {{\bf{g}}_i}{x_i}}}_{\text{inter-pair interference}}  + \underbrace{{\bf{g}}_{k'}^T{\bf{F}}{{\bf{n}}_{\rm{r}}}}_{\text{noise from relay}} + \underbrace{{n_{k'}}}_\text{noise at user}
\end{split}
\end{equation}
where $(k,k')$ is defined to indicate the $\left\lceil {k/2} \right\rceil$th\footnote{It is clear that $\left\lceil {k/2} \right\rceil=\left\lceil {k'/2} \right\rceil$.} user pair, and $n_{k'}$ represents the AWGN noise at the $k'$th user side with zero mean and variance of $\sigma_n^2$.

Using the estimated CSI, the relay calculates and sends out the SIC coefficient ${\hat{\bf g}}_{k}^T{\bf{F}}{{\hat {\bf{g}}}_{k}}$~($k\in\{1,\cdots,2K\}$) for each user. Hence, the received signal at the $k'$th user after SIC is rewritten as
\begin{equation}\label{rereceive_signal}
\begin{split}
&{{\tilde y}_{k'}} = \underbrace{\sqrt {{p_k}} {\bf{g}}_{k'}^T{\bf{F}}{{\bf{g}}_k}{x_k}}_{\text{desired signal}} + \underbrace{\sqrt {{p_{k'}}} \lambda_{k'}{x_{k'}}}_\text{residual self-interference} \\
&~~~+ \underbrace{{\bf{g}}_{k'}^T{\bf{F}}\sum\limits_{i \ne k,k'}^{2K} {\sqrt {{p_i}} {{\bf{g}}_i}{x_i}}}_{\text{inter-pair interference}}  + \underbrace{{\bf{g}}_{k'}^T{\bf{F}}{{\bf{n}}_{\rm{r}}}}_{\text{noise from relay}} + \underbrace{{n_{k'}}}_\text{noise at user}.
\end{split}
\end{equation}
where the residual self-interference involves $\lambda_{k'}={{\bf{g}}_{k'}^T{\bf{F}}{{\bf{g}}_{k'}} - {\hat{\bf g}}_{k'}^T{\bf{F}}{{\hat{\bf g}}_{k'}}}$, since the SIC coefficient ${\hat{\bf g}}_{k'}^T{\bf{F}}{{\hat {\bf{g}}}_{k'}}$ for user $k'$ is obtained from the estimated CSI. Here, we suppose that there is no error during the SIC coefficients transmission from the relay.

\subsection{MRC/MRT Processing}
In this subsection, the simple and widely used MRC/MRT beamforming is adopted. According to \cite{Liang2008}, the imperfect CSI based MRC/MRT beamforming is given by
\begin{equation}\label{F}
{\bf{F}} = \alpha_1 {{\hat{\bf G}}^*}{\bf{T}}{{\hat{\bf G}}^H}
\end{equation}
where ${\bf{T}} = diag\left\{ {{{\bf{T}}_1},{{\bf{T}}_2}, \ldots {{\bf{T}}_K}} \right\}$ is the block diagonal permutation matrix indicating the user pairing format with ${{\bf{T}}_1} = {{\bf{T}}_2} =  \cdots  = {{\bf{T}}_K} = [0~1;~1~0]$, and $\alpha_1$ is a normalization constant, chosen to satisfy the power constraint at the relay station in (\ref{power_consR}).

By substituting (\ref{F}) into (\ref{power_consR}), we have
\begin{equation}\label{alpha}
\begin{split}
&\alpha_1 \mathop  = \limits^{(a)} \sqrt {\frac{{{P_{\rm{R}}}}}{{{\rm E}\left[ {{{\left\| {{{\hat{\bf G}}^*}{\bf{T}}{{\hat{\bf G}}^H}{\tilde{\bf G}}{\bf x}} \right\|}^2}} \right] + {\rm E}\left[ {{{\left\| {{{\hat{\bf G}}^*}{\bf{T}}{{\hat{\bf G}}^H}{{\bf{n}}_{\rm{r}}}} \right\|}^2}} \right]}}}\\
& \mathop  = \limits^{(b)}\sqrt {\frac{{{P_{\rm{R}}}}}{{N\left( {N + 1} \right)\left[ {2\left( {\Psi  + \sigma _n^2} \right)\hat\Phi  + \left( {N + 1} \right)\sum\limits_{i = 1}^K {{\hat\psi _i}{\hat\phi _i}}} \right]}}}
\end{split}
\end{equation}
where
\begin{equation}\label{alpha_ext}
\begin{split}
&\hat\Phi  = \sum\limits_{i = 1}^K {\hat\phi _i},{\hat\phi _i} = \hat\sigma _{2i - 1}^2\hat\sigma _{2i}^2,{\hat\psi _i} = {p_{2i - 1}}\hat\sigma _{2i - 1}^2 + {p_{2i}}\hat\sigma _{2i}^2\\
&~~~~~\Psi  = \sum\limits_{i = 1}^K {\left( {{p_{2i - 1}}{\sigma} _{2i - 1}^2 + {p_{2i}}{\sigma} _{2i}^2} \right)}.
\end{split}
\end{equation}
The detailed derivation of the equation is given in Appendix A.

\subsection{ZFR/ZFT Processing}\label{ZFRZFT}
When employing ZFR/ZFT with imperfect CSI, in which the pseudo-inverse of the estimated channels in (\ref{estimated_CSI_exp}) are needed for processing, the linear beamforming is given by~\cite{Liang2008}
\begin{equation}\label{hat_F}
{\bf{F}} = \alpha_2 {{{\hat {\bf G}}}^*}{\left( {{{{\hat {\bf G}}}^T}{{{\hat {\bf G}}}^*}} \right)^{ - 1}}{\bf{T}}{\left( {{{{\hat {\bf G}}}^H}{\hat {\bf G}}} \right)^{ - 1}}{{\hat {\bf G}}^H} = \alpha_2 {\hat {\bar {\bf{G}}}}^*{\bf{T}}{\hat {\bar {\bf{G}}}}^H
\end{equation}
where ${\hat {\bar {\bf{G}}}} = {\hat {\bf G}}{\left( {{{{\hat {\bf G}}}^H}{\hat {\bf G}}} \right)^{ - 1}}$ and ${\alpha_2}$ is the normalization constant, chosen to satisfy the transmit power constraints at the relay. Notably, SIC is not necessary as ZFR/ZFT leads to ${\hat{\bf g}}_{k'}^T{\bf{F}}{{\hat {\bf{g}}}_{k'}}=0$~($\forall k'\in\{1,\cdots,2K\}$). On the basis of (\ref{hat_F}) and ${\rm Tr}\left\{ {{\bf{AB}}} \right\} = {\rm Tr}\left\{ {{\bf{BA}}} \right\}$, we have
\begin{equation}\label{hat_alpha}
\begin{split}
\alpha_2 & \mathop  = \limits^{(a)} \sqrt {\frac{{{P_{\rm{R}}}}}{{{\rm E}\left[ {{{\left\| {{\hat {\bar {\bf{G}}}}^*{\bf{T}}{\hat {\bar {\bf{G}}}}^H\tilde{\bf{G}}{\bf x}} \right\|}^2}} \right] + {\rm E}\left[ {{{\left\| {{\hat {\bar {\bf{G}}}}^*{\bf{T}}{\hat {\bar {\bf{G}}}}^H{{\bf{n}}_{\rm{r}}}} \right\|}^2}} \right]}}}\\
& \mathop  = \limits^{(b)} \sqrt {\frac{{{P_{\rm{R}}}}}{{\sum\limits_{i = 1}^{2K} {\frac{{{p_{i'}}}}{{\left( {N - 2K - 1} \right)\hat \sigma _i^2}} + \hat \eta \left( {\sum\limits_{j = 1}^{2K} {{p_j}\sigma _{{\xi _j}}^2}  + \sigma _n^2} \right)} }}}
\end{split}
\end{equation}
where $\hat \eta  = \sum\limits_{j = 1}^{2K} {\frac{1}{{\left( {N - 2K} \right)\left( {N - 2K - 3} \right)\hat \sigma _j^2\hat \sigma _{j'}^2}}}$. The detailed derivation of (\ref{hat_alpha}) is given in Appendix B.

\section{Achievable Rate Analysis}\label{Achiev_Rate}
In this section, a general form of the ergodic achievable rate of the transmission link $k\to k'$ for MRC/MRT processing is given first, followed by a rate expression for ZFR/ZFT. In order to obtain a basic and insightful expression that can be used for power allocation optimization, a simplified capacity lower bound is derived utilizing the technique of \cite{Marzetta2006,Jose2011}, in which the received signal is rewritten as a known mean times the desired symbol, plus an uncorrelated effective noise. The worst-case uncorrelated effective noise, where each additive term is treated as independent Gaussian noise of the same variance, is employed to derive a lower bound.

From (\ref{rereceive_signal}), the ergodic achievable rate of the transmission link $k\to k'$ is expressed as (\ref{ICSI_rate}).
\begin{table*}
\begin{equation}\label{ICSI_rate}
\begin{split}
\gamma _{k'}^{{\rm{ICSI}}} = {\rm E}\left[{\log _2}\left( {1 + \frac{{{p_k}{{\left| {{\bf{g}}_{k'}^T{\bf{F}}{{\bf{g}}_k}} \right|}^2}}}{{{{p_{k'}}{{\left| \lambda_{k'}\right|}^2}}+\sum\limits_{i \ne k,k'}^{2K} {{p_i}\left| {{\bf{g}}_{k'}^T{\bf{F}}{{\bf{g}}_i}} \right|^2 + {{\left\| {{\bf{g}}_{k'}^T{\bf{F}}} \right\|}^2}\sigma _n^2 + \sigma _n^2} }}} \right)\right].
\end{split}
\end{equation}
\end{table*}

\emph{Remark 1:} Here, the ergodic achievable rate is valid based on the assumption that the receiving user $k'$ knows perfectly ${{\bf{g}}_{k'}^T{\bf{F}}{{\bf{g}}_k}}$ in the detection process. To demonstrate the accuracy of the derived lower bounds, we compare the lower bounds with Monte-Carlo realized (\ref{ICSI_rate}) in Section \ref{NumericalResults}. The normalization constant for $\bf F$ in (\ref{ICSI_rate}) is assumed to be calculated based on instantaneous CSI by satisfying ${P_{\rm{R}}} = {{{\left\| \hat {\bf{r}} \right\|}^2}}$.

Further derivation of (\ref{ICSI_rate}) is difficult because of the intractability to carry out the ensemble average analytically. Instead, we adopt the technique in \cite{Marzetta2006} to derive a worst-case lower bound of the achievable rate. The first step is to rewrite $\sqrt {{p_k}}{\bf{g}}_{k'}^T{\bf{F}}{{\bf{g}}_k}{x_k}$ in (\ref{rereceive_signal}) as the sum of $\sqrt {{p_k}}{\rm E}\left[ {\bf{g}}_{k'}^T{\bf{F}}{{\bf{g}}_k}\right]{x_k}$ and $\sqrt {{p_k}} \left( {{\bf{g}}_{k'}^T{\bf{F}}{{\bf{g}}_k} - {\rm E}\left[ {{\bf{g}}_{k'}^T{\bf{F}}{{\bf{g}}_k}} \right]} \right){x_k}$, where the first part is now considered as the ``desired signal''. That is, (\ref{rereceive_signal}) can be expressed as
\begin{equation}\label{app_receive_signal}
{{\tilde y}_{k'}} = \underbrace{\sqrt {{p_k}} {\rm E}\left[ {{\bf{g}}_{k'}^T{\bf{F}}{{\bf{g}}_k}} \right]{x_k}}_{\text{desired signal}} + \underbrace{{{\tilde n}_{k'}}}_{\text{effective noise}}
\end{equation}
where ${{\tilde n}_{k'}}$ is considered as the effective noise and given by
\begin{equation}\label{eff_noise}
\begin{split}
{{\tilde n}_{k'}} \buildrel \Delta \over = &\sqrt {{p_k}} \left( {{\bf{g}}_{k'}^T{\bf{F}}{{\bf{g}}_k} - {\rm E}\left[ {{\bf{g}}_{k'}^T{\bf{F}}{{\bf{g}}_k}} \right]} \right){x_k} + \sqrt {{p_{k'}}} \lambda_{k'}{x_{k'}} \\
&+ {\bf{g}}_{k'}^T{\bf{F}}\sum\limits_{i \ne k,k'}^{2K} {\sqrt {{p_i}} {{\bf{g}}_i}{x_i}}  + {\bf{g}}_{k'}^T{\bf{F}}{{\bf{n}}_{\rm{r}}} + {n_{k'}}.
\end{split}
\end{equation}
It is straightforward to show that the first term ``desired signal'' and the second term ``effective noise'' in (\ref{app_receive_signal}) are uncorrelated. The exact pdf of ${\tilde n}_{k'}$ is not easy to obtain, but we know that the worst-case is to approximate the effective noise as independently Gaussian distributed~\cite{Marzetta2006}. Since the relay is equipped with large-scale antenna arrays by assuming $N\gg 2K\gg 1$, the central limit theorem provides a tight statistical CSI based lower bound for the achievable rate. Then, the statistical CSI based achievable rate lower bound of the transmission link $k\to k'$ can be obtained as
\begin{equation}\label{SCSI_rate}
\begin{split}
&\gamma _{k'}^{{\rm{SCSI}}} = \\
&{\log _2}\left( {1 + \frac{{{p_k}{{\left| {{\rm E}\left[ {{\bf{g}}_{k'}^T{\bf{F}}{{\bf{g}}_k}} \right]} \right|}^2}}}{{{p_k}{\rm Var}\left[ {{\bf{g}}_{k'}^T{\bf{F}}{{\bf{g}}_k}} \right] + {\rm{S}}{{\rm{I}}_{k'}}+ {\rm{I}}{{\rm{P}}_{k'}} + {\rm{N}}{{\rm{R}}_{k'}} + {\rm{N}}{{\rm{U}}_{k'}}}}} \right)
\end{split}
\end{equation}
where ${\rm{S}}{{\rm{I}}_{k'}}$, ${\rm{I}}{{\rm{P}}_{k'}}$, ${\rm{N}}{{\rm{R}}_{k'}}$ and ${\rm{N}}{{\rm{U}}_{k'}}$ denote the residual self-interference after SIC, the inter-pair interference, the amplified noise from relay and the noise at user, respectively, i.e.,
\begin{subequations}
    \label{SCSI_rate_ext}
    \begin{align}
    \begin{split}
    {\rm{S}}{{\rm{I}}_{k'}} \buildrel \Delta \over = &p_{k'}{\rm E}\left[ {{{\left| {{\bf{g}}_{k'}^T{\bf{F}}{{\bf{g}}_{k'}} - {\hat{\bf g}}_{k'}^T{\bf{F}}{{ {\hat{\bf g}}}_{k'}}} \right|}^2}} \right]
    \end{split}\\
    \begin{split}
    {\rm{I}}{{\rm{P}}_{k'}} \buildrel \Delta \over = &\sum\limits_{i \ne k,k'}^{2K} {{p_i}{\rm E}\left[ {{{\left| {{\bf{g}}_{k'}^T{\bf{F}}{{\bf{g}}_i}} \right|}^2}} \right]}
    \end{split}\\
    \begin{split}
    {\rm{N}}{{\rm{R}}_{k'}} \buildrel \Delta \over = &{\rm E}\left[ {{{\left| {{\bf{g}}_{k'}^T{\bf{F}}{{\bf{n}}_{\rm{r}}}} \right|}^2}} \right],~{\rm{N}}{{\rm{U}}_{k'}} \buildrel \Delta \over = {\rm E}\left[ {{{\left| {{n_{k'}}} \right|}^2}} \right].
    \end{split}
    \end{align}
\end{subequations}
%

When MRC/MRT beamforming is employed, further mathematical derivation of (\ref{SCSI_rate}) leads to the following theorem:

\emph{Theorem 1:} With imperfect CSI based MRC/MRT, the ergodic achievable rate of the transmission link $k\to k'$, for a finite number of antennas at the relay, is lower bounded by (\ref{SCSI_rate_CF}),
\begin{table*}
\begin{equation}\label{SCSI_rate_CF}
{\text {MRC/MRT}}:~\gamma _{k'}^{{\rm{SCSI}}} = {\log _2}\left( {1 + {{{a_{k'}}{p_k}} \over {\sum\limits_{i = 1}^{2K} {\left( {b_{k',i}^{(1)} + b_{k',i}^{(2)}P_{\rm{R}}^{ - 1}} \right){p_i}}  + {c_{k'}}{p_{k'}} + \left( {d_{k'}^{(1)} + d_{k'}^{(2)}P_{\rm{R}}^{ - 1}} \right)}}} \right)
\end{equation}
\begin{equation}\label{hat_SCSI_rate_CF}
{\text {ZFR/ZFT}}:~\gamma _{k'}^{{\rm{SCSI}}} = {\log _2}\left( {1 + {{{e_{k'}}{p_k}} \over {\sum\limits_{i = 1}^{2K} {\left( {f_{k',i}^{(1)} + f_{k',i}^{(2)}P_{\rm{R}}^{ - 1}} \right){p_i}}  + {m_{k'}}{p_{k'}} + \left( {n_{k'}^{(1)} + n_{k'}^{(2)}P_{\rm{R}}^{ - 1}} \right)}}} \right)
\end{equation}
\end{table*}
where ${a_{k'}} = N\left( {N + 1} \right)\hat\sigma _k^4\hat\sigma _{k'}^4$, $b_{k',i}^{(1)} = \left( {N + 1} \right)\left( {\sigma _i^2\hat \sigma _{k'}^4\hat \sigma _k^2 + \sigma _{k'}^2\hat \sigma _i^4\hat \sigma _{i'}^2} \right) + 2\sigma _i^2\sigma _{k'}^2\hat \Phi $, $b_{k',i}^{(2)} = \sigma _n^2\left[ 2\hat \Phi \sigma _i^2\sigma _{i'}^2 \right.$ $\left.+ \left( {N + 1} \right)\hat \sigma _i^4\hat \sigma _{i'}^2 \right]$, ${c_{k'}} = 2\left[ \left( {N + 1} \right)\left(\sigma_{k'}^2-\right.\right.$ $\left.\left.2\hat\sigma_{k'}^2\right)\hat\sigma _k^2\hat\sigma _{k'}^4 + \left(\sigma_{k'}^4 -2\hat\sigma_{k'}^4\right)\hat\Phi \right]$, $d_{k'}^{(1)} = \sigma _n^2\left[ \left( {N + 1} \right)\hat \sigma _k^2\hat \sigma _{k'}^4 \right.$ $\left.+ 2\sigma _{k'}^2\hat \Phi  \right]$, and $d_{k'}^{(2)} = 2\sigma _n^4\hat \Phi $.

\emph{Proof:} See Appendix C.

For imperfect CSI based ZFR/ZFT processing, a closed-form expression for the achievable rate in (\ref{SCSI_rate}) is derived as follows:

\emph{Theorem 2:} With imperfect CSI based ZFR/ZFT beamforming, the achievable rate of the transmission link $k\to k'$, for a finite number of antennas at the relay, is lower bounded by (\ref{hat_SCSI_rate_CF}),
where ${{e}_{k'}} = {{e}_k} = 1$, $f_{k',i}^{(1)} = {{\sigma _{{\xi _i}}^2} \over {\left( {N - 2K - 1} \right)\hat \sigma _k^2}} + {{\sigma _{{\xi _{k'}}}^2} \over {\left( {N - 2K - 1} \right)\hat \sigma _{i'}^2}} + \sigma _{{\xi _{k'}}}^2\sigma _{{\xi _i}}^2\hat \eta$, $f_{k',i}^{(2)} = \sigma _n^2\left( {{1 \over {\left( {N - 2K - 1} \right)\hat \sigma _{i'}^2}} + \sigma _{{\xi _i}}^2\hat \eta } \right)$, ${{m}_{k'}} = \sigma _{{\xi _{k'}}}^4\hat \eta$, $n_{k'}^{(1)} = {{\sigma _n^2} \over {\left( {N - 2K - 1} \right)\hat \sigma _k^2}} + \sigma _n^2\sigma _{{\xi _{k'}}}^2\hat \eta$, and $n_{k'}^{(2)} = \sigma _n^4\hat \eta $.

\emph{Proof:} See Appendix D.

Theorems 1 and 2 are also valid for conventional MIMO systems, while the bounds become less tight as the antenna scale goes down. The capacity lower bounds for perfect CSI can always be obtained by setting $\sigma_{\xi_k}^2= 0$ and $\hat\sigma_{k}^2 = \sigma_{k}^2$~($k \in \{1,\cdots,2K\}$) in (\ref{SCSI_rate_CF}) and (\ref{hat_SCSI_rate_CF}). Moreover, it can be observed from (\ref{SCSI_rate_CF}) that when the estimation error is severe, the residual SI occupies the major part of the imperfect CSI effect in comparison to other terms. On the other hand, if channel estimation is rather accurate, the residual SI has slight effects in comparison to other terms. While for ZFR/ZFT, both the residual SI and inter-pair interference are determined by the channel estimation accuracy.

\subsection{Asymptotic Analysis with Massive Arrays}
Based on the derived closed-form expressions for the achievable rate in (\ref{SCSI_rate_CF}) and (\ref{hat_SCSI_rate_CF}), this section provides the asymptotic analysis under two different cases when the number of relay antennas approaches to infinity. Suppose that all users have the same transmit power, i.e., ${p_1} = {p_2}  \cdots  = {p_{2K}} = P_{\text S}$.

\emph{Proposition 1:} In case I where $p_{\rm P}$ is fixed, ${p_i} = P_{\text S} = \frac{{{E_{\rm{S}}}}}{N^\rho}$~($i = 1,2,\cdots,2K$), ${P_{\rm{R}}} = \frac{{{E_{\rm{R}}}}}{N^\theta}$, and $E_{\rm{S}}$ and $E_{\rm{R}}$ are fixed, to achieve non-vanishing user rate as $N \to \infty$, the user and relay transmit power scaling factor $\rho$ and $\theta$ must satisfy $0\le \rho \le 1$ and $0\le \theta \le 1$. When $\rho = 1$ and $\theta = 1$, the asymptotic achievable rate expressions of the transmission link $k\to k'$ for imperfect CSI based MRC/MRT and ZFR/ZFT are (\ref{SCSI_rate_P1}) and (\ref{hat_SCSI_rate_P1}),
\begin{table*}
\begin{equation}\label{SCSI_rate_P1}
{\text {MRC/MRT}}:~\gamma _{k'}^{{\rm{SCSI}}}\mathop  {\longrightarrow} \limits_{N \to \infty }^{a.s.} {\log _2}\left( {1 + \frac{{{E_{\rm{S}}}{E_{\rm{R}}}\hat\sigma _k^4\hat\sigma _{k'}^4}}{{{E_{\rm{S}}}\sigma _n^2\sum\limits_{i = 1}^{2K} {\hat\sigma _i^4\hat\sigma _{i'}^2}  + {E_{\rm{R}}}\sigma _n^2\hat\sigma _k^2\hat\sigma _{k'}^4 + 2\hat\Phi \sigma _n^4}}} \right)
\end{equation}
\begin{equation}\label{hat_SCSI_rate_P1}
{\text {ZFR/ZFT}}:~\gamma _{k'}^{{\rm{SCSI}}}\mathop  {\longrightarrow} \limits_{N \to \infty }^{a.s.} {\log _2}\left( {1 + \frac{{{E_{\rm{R}}}{E_{\rm{S}}}\hat \sigma _k^2}}{{{E_{\rm{R}}}\sigma _n^2 + \hat \sigma _k^2\sigma _n^2\sum\limits_{i = 1}^{2K} {\left( {\frac{{{E_{\rm{S}}}}}{{\hat \sigma _i^2}} + \frac{{\sigma _n^2}}{{\hat \sigma _i^2\hat \sigma _{i'}^2}}} \right)} }}} \right)
\end{equation}
\begin{equation}\label{SCSI_rate_P2}
{\text {MRC/MRT}}:~\gamma _{k'}^{{\rm{SCSI}}}\mathop  {\longrightarrow} \limits_{N \to \infty }^{a.s.} {\log _2}\left( {1 + \frac{{{\tau ^2}E_{\rm P}^2E_{\rm{S}}{E_{\rm{R}}}\sigma _k^8\sigma _{k'}^8}}{{\tau E_{\rm P}\sigma _n^4\left[E_{\rm{S}}\sum\limits_{i = 1}^{2K} {\sigma _i^8\sigma _{i'}^4}  + {E_{\rm{R}}}\sigma _k^4\sigma _{k'}^8 \right]+ 2\sigma _n^8\sum\limits_{j = 1}^K {\sigma _j^4\sigma _{j'}^4} }}} \right)
\end{equation}
\begin{equation}\label{hat_SCSI_rate_P2}
{\text {ZFR/ZFT}}:~\gamma _{k'}^{{\rm{SCSI}}}\mathop  {\longrightarrow} \limits_{N \to \infty }^{a.s.} {\log _2}\left( {1 + \frac{{{\tau ^2}E_{\rm P}^2E_{\rm{S}}{E_{\rm{R}}}\sigma _k^4}}{{\tau E_{\rm P}{E_{\rm{R}}}\sigma _n^4 + \sigma _k^4\sigma _n^4\sum\limits_{i = 1}^{2K} {\left( {\frac{{\tau E_{\rm P}E_{\rm{S}}}}{{\sigma _i^4}} + \frac{{\sigma _n^4}}{{\sigma _i^4\sigma _{i'}^4}}} \right)} }}} \right)
\end{equation}
\end{table*}
respectively, which show that the transmit powers at both users and relay sides can be scaled down by up to $\frac{1}{N}$ to maintain a given rate in case I. When $\rho < 1$ and $\theta < 1$, the asymptotic achievable rate of each user approaches to infinity as $N \to \infty$.

In case II where ${p_{\rm P}} = \frac{{{E_{\rm P}}}}{N^\varsigma}$, ${p_i} = P_{\text S} = \frac{{{E_{\rm{S}}}}}{N^\rho}$~($i = 1,2,\cdots,2K$), ${P_{\rm{R}}} = \frac{{{E_{\rm{R}}}}}{N^\theta}$, and $E_{\rm{S}}$ and $E_{\rm{R}}$ are fixed, to achieve non-vanishing user rate as $N \to \infty$, the pilot, user and relay transmit power scaling factors $\varsigma$, $\rho$ and $\theta$ must satisfy $0<\varsigma\le 1$, $0\le \rho \le 1 - \varsigma$ and $0\le \theta \le 1- \varsigma$. When $0\le\varsigma<1$, $\rho = 1- \varsigma$ and $\theta = 1- \varsigma$, the asymptotic achievable rate of the transmission link $k\to k'$ for imperfect CSI based MRC/MRT and ZFR/ZFT are (\ref{SCSI_rate_P2}) and (\ref{hat_SCSI_rate_P2}), respectively, from which we conclude that the transmit powers of each user and the relay can only be reduced by up to $\frac{1}{N^{1-\varsigma}}$ when the pilot transmit power is set as $p_{\rm P}= \frac{{{E_{\rm P}}}}{N^\varsigma}$, in order to maintain a given spectral efficiency. Similarly, when $\rho < 1- \varsigma$ and $\theta < 1- \varsigma$, the asymptotic achievable rate of each user approaches to infinity as $N \to \infty$.

\emph{Remark 2:} When the pilot power scaling factor $\varsigma=1$, which means that the pilot power scales down by $\frac{1}{N}$, to guarantee user rate there is $\rho=\theta=0$, which means the relay and user transmit power must stay constant and do not scale down with $N$. The achievable rate in this case can be derived from (\ref{SCSI_rate_CF}) and (\ref{hat_SCSI_rate_CF}), but not shown here due to space limitation. It is found that channel estimation error induced interference and inter-pair interference cannot be eliminated when $p_{\rm P}$ is scaled down proportionally to $\frac{1}{N}$ with fixed ${p_i}$ and ${P_R}$ in case II.

It can be observed from Theorems 1 and 2 that for fixed $\sigma_i$~($i = 1,\cdots,2K$), $\sigma _n$, and $p_{\rm P}$, the achievable rate of each pair-wise user transmission link depends on the user power, i.e., the values of $p_i$ ($i=1,\cdots,2K$), and the relay power. Next we propose the optimal power allocation for the studied system.

\section{Power Allocation Schemes}\label{PA}
In this section, a power allocation problem is first formulated and solved for multi-pair users in the MA phase transmission and the relay in the BC phase, which maximizes the sum spectral efficiency\footnote{The objective of power allocation can also be minimizing the total power consumption or maximizing the minimum achievable rate, which can be formulated and solved using the similar method.}. The achievable rate of a transmission link $\gamma_k$ for $k\in \{1,\cdots,2K\}$ used in the optimization refers to the SCSI based achievable rate, given in (\ref{SCSI_rate_CF}) and (\ref{hat_SCSI_rate_CF}). Power allocation can be performed at the relay side according to the SCSI, and then the relay notifies the user pairs their allocated power values. Moreover, closed-form asymptotic power allocation solutions are presented for MRC/MRT and ZFR/ZFT, respectively, for the asymptotic regimes with high SNR and $N \to \infty $.

\subsection{Optimal Power Allocation (OPA)}
Most power optimization in communications aims to maximize the sum spectral efficiency, which is defined as the sum-rate (in bits) per channel use. Assuming that $T$ is the length of the coherent interval (in symbols), in which $\tau$ symbols are used for channel estimation, the sum spectral efficiency\footnote{Here, the loss due to relay sending out pairwise effective channel coefficients and SIC coefficients is taken into account. The transmission is assumed to be perfect and the overhead is 2 (symbols) for the pairwise effective channel coefficient and SIC coefficient per user.} denoted as $S \buildrel \Delta \over = \frac{{T - \tau - 2 }}{T}\sum\limits_{k = 1}^{2K} {{\gamma _k}} = \frac{{T - \tau - 2 }}{T}{\log _2}\prod\limits_{k = 1}^{2K} {\left( {1 + {\chi _k}} \right)}$, where $\chi _k={{{a_{k}}{p_{k'}}} \over {\sum\limits_{i = 1}^{2K} {\left( {b_{k,i}^{(1)} + b_{k,i}^{(2)}P_{\rm{R}}^{ - 1}} \right){p_i}}  + {c_{k}}{p_{k}} + \left( {d_{k}^{(1)} + d_{k}^{(2)}P_{\rm{R}}^{ - 1}} \right)}}$ is the signal-to-interference-plus-noise ratio (SINR). Then, the power allocation problem to maximize the sum spectral efficiency can be formulated as
\begin{subequations}
    \label{Opt3}
    \begin{align}
    \label{objective3}
    \begin{split}
    \mathop {\max }\limits_{{p_i,P_{\rm{R}}}}~&\frac{{T - \tau - 2}}{T}{\log _2}\prod\limits_{k = 1}^{2K} {\left( {1 + {\chi _k}} \right)}
    \end{split}\\
    \label{constraints31}
    \begin{split}
    {\rm{s.t.}}~~& \sum\limits_{i = 1}^{2K} {{p_i}} + P_{\rm{R}}  \le P
    \end{split}\\
    \label{constraints32}
    \begin{split}
    & 0 \le {p_{k}} \le {P_0},~0 \le P_{\rm{R}} \le {P_{\rm{R,0}}},~k=1,\cdots,2K
    \end{split}
    \end{align}
\end{subequations}
where $P$ in constraint (\ref{constraints31}) is the total power allocated to all the users and relay, and constraints (\ref{constraints32}) specify the peak power limits $P_0$ and ${P_{\rm{R,0}}}$ for each user $k$ and relay, respectively. The objective in (\ref{objective3}) can be equivalently rewritten as $\mathop {\min }\limits_{{p_i},P_{\rm{R}}}~ \left[\prod\limits_{k = 1}^{2K} {\left( {1 + {\chi _k}} \right)}\right]^{-1}$, as ${\log _2}(x)$ is a monotonic increasing function of $x$. We can see that the constraints are posynomial functions. If the objective function is a monomial or posynomial function, the problem (\ref{Opt3}) becomes a GP which can be reformulated as a convex problem, and thus, can be solved efficiently by convex optimization tools, such as CVX~\cite{cvx2008}. However, the rewritten objective function for (\ref{Opt3}) is still neither a monomial nor posynomial, making solving the problem directly by the convex optimization tools impossible. To solve this problem, an approximation for the objective function can be efficiently found by using the technique in \cite{Weeraddana2011}. Specifically, according to \cite[Lemma 1]{Weeraddana2011}, we can use a monomial function ${{\kappa _k}\chi _k^{{\eta _k}}}$ to approximate $({1 + {\chi _k}})$ near an arbitrary point ${{\hat \chi }_k}>0$, where ${\eta _k} \buildrel \Delta \over = {{\hat \chi }_k}{\left( {1 + {{\hat \chi }_k}} \right)^{ - 1}}$ and ${\kappa _k} \buildrel \Delta \over = \hat \chi _k^{ - {\eta _k}}\left( {1 + {{\hat \chi }_k}} \right)$.
Consequently, the objective function can be approximated as $\prod\limits_{k = 1}^{2K} {\left( {1 + {\chi _k}} \right)}  \approx \prod\limits_{k = 1}^{2K} {{\kappa _k}\chi _k^{{\eta _k}}}$, which is a monomial function. In this way, the problem is transformed into a GP problem by the approximation.

Similar to \cite{Weeraddana2011}, a successive approximation algorithm for the power allocation problem in (\ref{Opt3}) is proposed as Algorithm 1. (\ref{Trans2_Opt3_Cons1}) is the relaxed constraint. Notably, the parameter $\beta$ here is utilized to control the desired approximation accuracy. The accuracy is high when $\beta$ is close to 1, but the convergence rate is low, and vice versa. As shown in \cite{Weeraddana2011}, $\beta=1.1$ is an option that introduces a good accuracy trade off in most practical cases.
\vspace*{2em}
\begin{table*}[htbp]
\vspace{0em}
\label{tab:1}       
\centering
\begin{tabular}{p{0.9\textwidth}}
\toprule
\textbf{Algorithm 1:} Successive approximation algorithm for power allocation \\
\midrule
\textbf{Initialization:} Given tolerance $\varepsilon >0$, the maximum number of iterations $L$, and parameter $\beta >1$. Set $m = 1$. Select the initial values $\chi_{k,1}$ for $\chi_k$~($k=1,2,\cdots,2K$).\\
\textbf{Repeat:}
    \begin{enumerate}
    \item Compute ${\eta _{k,m}}  = {\chi _{k,m}}{\left( {1 + {\chi _{k,m}}} \right)^{ - 1}}$ and ${\kappa _{k,m}} = \chi _{k,m}^{ - {\eta _{k,m}}}\left( {1 + {\chi _{k,m}}} \right)$;
    \item Solve the GP:

    {\begin{subequations}
    \label{Trans2_Opt3}
    \begin{align}
    \mathop {\min }\limits_{{p_i},P_{\rm R},\chi _{k}}~&\left[\prod\limits_{k = 1}^{2K} {{\kappa _{k,m}}\chi _k^{{\eta _{k,m}}}}\right]^{-1}\\
    \label{Trans2_Opt3_Cons1}
    {\rm{s.t.}}~~&{{{a_{k}}{p_{k'}}} \over {\sum\limits_{i = 1}^{2K} {\left( {b_{k,i}^{(1)} + b_{k,i}^{(2)}P_{\rm{R}}^{ - 1}} \right){p_i}}  + {c_{k}}{p_{k}} + \left( {d_{k}^{(1)} + d_{k}^{(2)}P_{\rm{R}}^{ - 1}} \right)}}\le \chi_k\\
    &\text{(\ref{constraints31})},~\text{(\ref{constraints32})},~{\beta ^{ - 1}}{\chi_{k,m}} \le {\chi _k} \le \beta {\chi _{k,m}}
    \end{align}
    \end{subequations}}

    \item Set $m = m+1$, and update $\chi _{k,m}=\chi _{k}^*$, where $\chi _{k}^*$ ~($k=1,2,\cdots,2K$) are obtained based on the solutions $p_i^*$~($i=1,2,\cdots,2K$) and $P_{\rm R}^*$ of the GP;
    \end{enumerate}
\textbf{Until:} Stop if $\mathop {\max }\limits_k \left| \frac{{\chi _{k,m}} - {\chi _{k,m-1}}}{\chi _{k,m}} \right| < \varepsilon$ or $m=L$;\\
\textbf{Output:} Output $p_i^*$~($i=1,2,\cdots,2K$) and $P_{\rm R}^*$ as the solutions.\\
\bottomrule
\end{tabular}
\end{table*}

\subsection{Asymptotically Optimal Power Allocation (AOPA)}
Obviously, the optimal power allocation scheme in Algorithm 1 is an iterative numerical solution with no closed-form. However, more tractable expressions can be found for MRC/MRT and ZFR/ZFT, respectively, when we consider the asymptotic regimes with high SNR and $N \to \infty $.

\subsubsection{AOPA for MRC/MRT}
Suppose that the SNRs at both the relay and user sides are very high, i.e., $P_{\rm R}\gg {\sigma_n^2}$, ${p_{\rm P}}\gg {\sigma_n^2}$ and ${p_i}\gg {\sigma_n^2}$~($i = 1,\cdots,2K$), and $N \gg K \to \infty$. Then the lower bound can be simplified as given by the following Lemma.

\emph{Lemma 1:} When $P_{\rm R}\gg {\sigma_n^2}$, ${p_{\rm P}}\gg {\sigma_n^2}$, ${p_i}\gg {\sigma_n^2}$~($i = 1,\cdots,2K$), and $N \gg K \to \infty$, the rate of the transmission link $k\to k'$ can be approximated as
\begin{equation}\label{SCSI_rate_CF_app}
\gamma _{k'}^{{\rm{SCSI}}} \approx {\log _2}\left( 1+ {{N\sigma _k^4\sigma _{k'}^2{p_k}} \over {\sum\limits_{i \ne k'}^{2K} {\left( {\sigma _i^2\sigma _{k'}^2\sigma _k^2 + \sigma _i^4\sigma _{i'}^2} \right){p_i}} }}\right).
\end{equation}

\emph{Proof:} Firstly, we have $\hat\sigma_i^2\approx \sigma_i^2$ for $i = 1,\cdots,2K$ due to ${p_{\rm P}}\gg {\sigma_n^2}$. Then, we divide both the denominator and numerator of the SINR in (\ref{SCSI_rate_CF}) by $(N+1)$. Each item with $\frac{1}{N+1}$ in the denominator is able to be ignored based on $N \to \infty$. Then, according to $\frac{\sigma_n^2}{P_{\rm R}}\approx 0$ and ${p_i}\gg {\sigma_n^2}$ for $i = 1,\cdots,2K$, (\ref{SCSI_rate_CF_app}) can be obtained.

In order to obtain a closed-form solution for asymptotically optimal power allocation, we set the fixed link condition $\sigma_i^2\sigma_{i'}^2=C$ for $i = 1,\cdots,2K$. Since the approximated rate expressions involve no $P_{\rm R}$, we do not need to find the optimal solution for $P_{\rm R}$. For a fixed $P_{\rm{R}}$, (\ref{Opt3}) can be rewritten as
\begin{subequations}
    \label{Opt3_app}
    \begin{align}
    \label{objective3_app}
    \begin{split}
    \mathop {\max }\limits_{{p_i}}~&\prod\limits_{k = 1}^{2K} {\left( {1 + {{N\sigma _{k'}^2{p_{k'}}} \over {2\sum\limits_{i \ne k}^{2K} {\sigma _i^2{p_i}} }}} \right)}
    \end{split}\\
    \label{constraints31_app}
    \begin{split}
    {\rm{s.t.}}~~& \sum\limits_{i = 1}^{2K} {{p_i}} \le P-P_{\rm{R}}
    \end{split}\\
    \label{constraints32_app}
    \begin{split}
    & 0 \le {p_{k}},~k=1,2,\cdots,2K.
    \end{split}
    \end{align}
\end{subequations}
where the peak power constraints for each user are ignored for analysis simplicity by assuming that the channel large-scale factors are on the same order of magnitude. Then, we have the following \emph{Theorem 3} with regard to the optimal allocated power for each user.

\emph{Theorem 3:} The optimal solution to (\ref{Opt3_app}) is obtained as
\begin{equation}\label{solution_app}
p_i^* = {{P-P_{\rm{R}}} \over {\sigma _i^2\sum\limits_{k = 1}^{2K} {{1 \over {\sigma _k^2}}} }}.
\end{equation}

\emph{Proof:} Since $\left|{{1 \over {2K - 1}}\sum\limits_{i \ne k'}^{2K} {\sigma _i^2{p_i}} -{1 \over {2K}}\sum\limits_{i = 1}^{2K} {\sigma _i^2{p_i}}}\right|\mathop  {\longrightarrow} \limits^{p} 0$~\cite{XianWu2015} in probability when $K \to \infty$, according to the inequality of arithmetic and geometric means, the objective in (\ref{objective3_app}) is upper bounded by
\begin{equation}\label{objective3_app_uppb}
\begin{split}
&\mathop \prod \limits_{k = 1}^{2K} {{\left( {2K - 1} \right)\sum\limits_{i = 1}^{2K} {\sigma _i^2{p_i}}  + NK\sigma _{k'}^2{p_{k'}}} \over {\left( {2K - 1} \right)\sum\limits_{i = 1}^{2K} {\sigma _i^2{p_i}} }} \\
\le & {\left( {\sum\limits_{k = 1}^{2K} {{{\left( {2K - 1} \right)\sum\limits_{i = 1}^{2K} {\sigma _i^2{p_i}}  + NK\sigma _{k'}^2{p_{k'}}} \over {2K\left( {2K - 1} \right)\sum\limits_{i = 1}^{2K} {\sigma _i^2{p_i}} }}} } \right)^{2K}}\\
= &{\left( {{1 \over {2K}} + {N \over {2\left( {2K - 1} \right)}}} \right)^{2K}}
\end{split}
\end{equation}
where the equality is achieved if and only if ${\sigma _i^2{p_i}} = A$ for $\forall i \in\{ 1,\cdots, 2K\}$. To maximize the sum rate, it is obvious that the total user power should reach the largest value $P-P_{\rm{R}}$ in (\ref{constraints31_app}). Thus (\ref{solution_app}) can be obtained to satisfy $\sum\limits_{i = 1}^{2K} {{p_i}} = P - P_{\rm{R}}$, indicating that the optimal allocated power for each user is inverse to its corresponding large-scale fading factor of the channel from the user to the relay, and proportional to the channel from the relay to its destination when MRC/MRT beamforming is used at the relay under the condition that the link end-to-end large-scale fading factors among all pairs are equal. Furthermore, the asymptotic sum-rate is independent of the allocated power.

\subsubsection{AOPA for ZFR/ZFT}
Similarly, we make the same assumption for ZFR/ZFT that the SNRs at both the relay and user sides are very high and $N \gg K \to \infty$. Then the following Lemma can be obtained.

\emph{Lemma 2:} When $P_{\rm R}\gg {\sigma_n^2}$, ${p_{\rm P}}\gg {\sigma_n^2}$, ${p_i}\gg {\sigma_n^2}$~($i = 1,\cdots,2K$), and $N \gg K \to \infty$, the rate of the transmission link $k\to k'$ can be approximated as
\begin{equation}\label{SCSI_rate_CF_app1}
\gamma _{k'}^{{\rm{SCSI}}} \approx {\log _2}\left( 1+ {{\left( {N - 2K - 1} \right)\sigma _k^2{p_k}} \over {\sigma _n^2}}\right).
\end{equation}
Therefore, without any assumptions on link conditions, (\ref{Opt3}) can be rewritten as
\begin{subequations}
    \label{Opt3_app1}
    \begin{align}
    \label{objective3_app1}
    \begin{split}
    \mathop {\max }\limits_{{p_i}}~&\prod\limits_{k = 1}^{2K} {\left( 1+ {{\left( {N - 2K - 1} \right)\sigma _{k'}^2{p_{k'}}} \over {\sigma _n^2}}\right)}
    \end{split}\\
    \label{constraints31_app1}
    \begin{split}
    {\rm{s.t.}}~~& (\ref{constraints31_app}), (\ref{constraints32_app})
    \end{split}
    \end{align}
\end{subequations}
\emph{Theorem 4} states the optimal allocated power for each user.

\emph{Theorem 4:} The optimal solution to (\ref{Opt3_app1}) is given by
\begin{equation}\label{solution_app1}
p_i^* = {\left[ {{1 \over \lambda } - {{\sigma _n^2} \over {\left( {N - 2K - 1} \right)\sigma _i^2}}} \right]^ + }
\end{equation}
where $\lambda  = {{2K} \mathord{\left/
 {\vphantom {{2K} {\left( {\sum\limits_{i = 1}^{2K} {{{\sigma _n^2} \over {\left( {N - 2K - 1} \right)\sigma _i^2}}}  + {P \over 2}} \right)}}} \right.
 \kern-\nulldelimiterspace} {\left( {\sum\limits_{i = 1}^{2K} {{{\sigma _n^2} \over {\left( {N - 2K - 1} \right)\sigma _i^2}}}  + {P \over 2}} \right)}}$ is chosen to satisfy (\ref{constraints31_app}).

\emph{Proof:} Obviously, we can obtain the solution to (\ref{Opt3_app1}) by the Lagrange multiplier approach associated with Karush-Kuhn-Tucker (KKT) conditions.

From (\ref{solution_app1}), it can be concluded that the allocated powers are equal for ZFR/ZFT when $N-2K\to \infty$.

\section{Numerical Results}\label{NumericalResults}
Simulations are conducted to validate the derived achievable rate expressions and examine the performance of the designed power allocation schemes, respectively. In the simulation study, we set the length of the coherent interval $T=200$ (symbols), the number of user pairs $K=10$, the training length $\tau=2K$, and the noise variance is normalized to be $\sigma_n^2=1$. Furthermore, ${\text {SNR}}={P_{\rm{R}}}$ is defined at the relay side.

\subsection{Validation of Achievable Rate Results}
\begin{figure}
   \centering
   \includegraphics[scale=0.5]{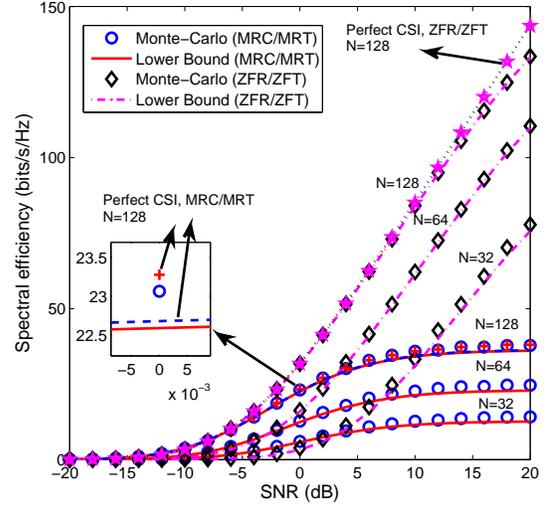}
   \caption{Spectral efficiency versus SNR for lower bounds and Monte-Carlo results ($p_{\rm P}=10$~dB, EPA, ${P_{\rm R}} ={2K}{P_{\rm S}}$).}
   \label{SCSI_ICSI1}
\end{figure}

Firstly, the effectiveness of the derived SCSI based achievable rate in (\ref{SCSI_rate_CF}) and (\ref{hat_SCSI_rate_CF}) is evaluated by comparing the spectral efficiency with the Monte-Carlo simulation results. For simplicity, we assume that the large-scale fading factors are $\sigma_i^2=1$ for all $i=1,2,\cdots,2K$ and equal power allocation for users is utilized with the total power $2KP_{\rm S}=P_{\text R}$. In Fig.~\ref{SCSI_ICSI1} with $p_{\rm P}=10$~dB, the spectral efficiency curves versus $P_{\rm S}$ obtained from the analytical lower bounds (\ref{SCSI_rate_CF}) and (\ref{hat_SCSI_rate_CF}), are compared with the ones given by the exact capacity expression (\ref{ICSI_rate}) obtained through Monte-Carlo simulation. It is evident that the relative performance gap between the capacity lower bound (\ref{hat_SCSI_rate_CF}) and the exact capacity (\ref{ICSI_rate}) for ZFR/ZFT is even smaller than that for MRC/MRT, especially with larger number of antennas. Moreover, Fig.~\ref{SCSI_ICSI1} shows that the spectral efficiency of ZFR/ZFT increases much faster than that of MRC/MRT as SNR increases. It is due to the fact that the effect of interference is much larger than that of the noise for higher SNR while ZFR/ZFT is able to null multi-user interference signals~\cite{Hong2013}. Hence, the effectiveness of the derived closed-form lower bounds for both MRC/MRT and ZFR/ZFT has been demonstrated.

\begin{figure}
   \centering
   \includegraphics[scale=0.5]{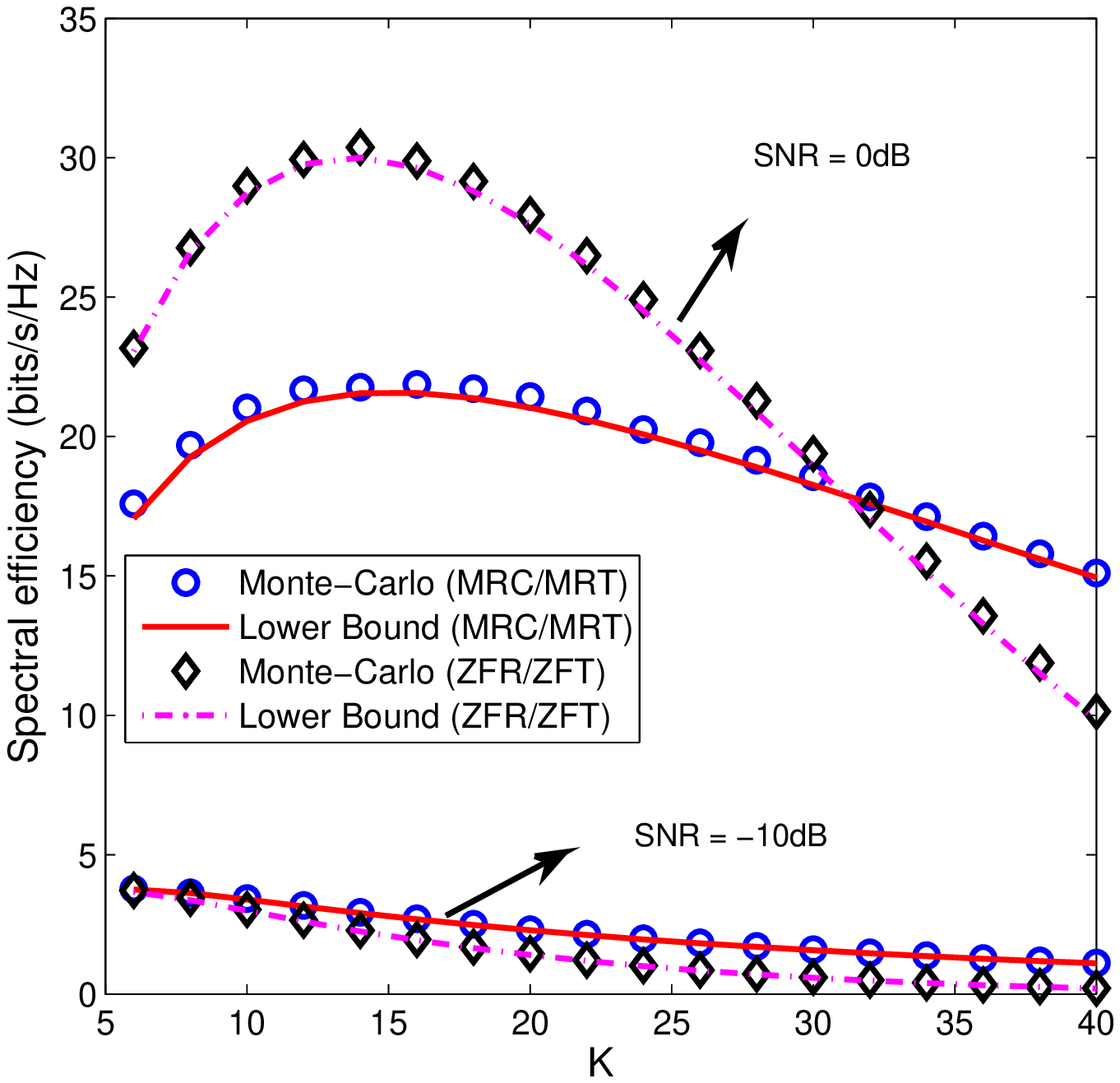}
   \caption{Spectral efficiency versus K for lower bounds and Monte-Carlo results ($p_{\rm P}=10$~dB, EPA, ${\rm SNR}= {P_{\rm R}}={2K}{P_{\rm S}}$, $N = 128$).}
   \label{SCSI_ICSI2}
\end{figure}

\begin{figure}
   \centering
   \includegraphics[scale=0.51]{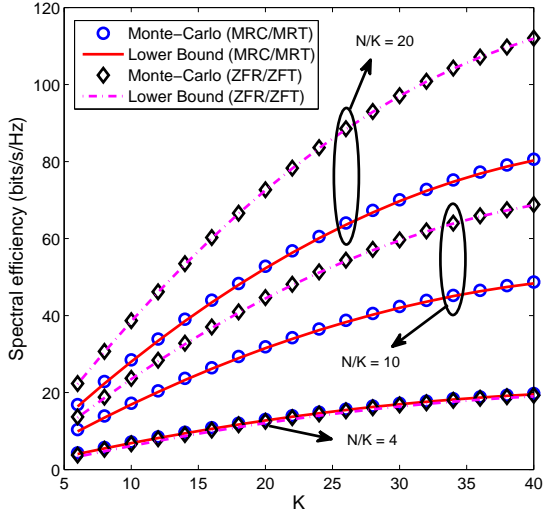}
   \caption{Spectral efficiency versus K for lower bounds and Monte-Carlo results ($p_{\rm P}=10$~dB, EPA, ${\rm SNR}= {P_{\rm R}}={2K}{P_{\rm S}}=0$~dB).}
   \label{SCSI_ICSI3}
\end{figure}

Fig.~\ref{SCSI_ICSI2} compares the spectral efficiency of MRC/MRT with that of ZFR/ZFT at different SNRs, which shows that ZFR/ZFT does not always outperform MRC/MRT in the massive MIMO two-way relaying systems. As $K$ increases, i.e., the average SNR ($\frac{P_{\rm{R}}}{2K}$) of each user decreases, the noise effect exceeds the interference effect causing wore performance of ZFR/ZFT. Fig.~\ref{SCSI_ICSI3} depicts the performance of ZFR/ZFT over MRC/MRT with different $N/K$ ratios. Different from the results obtained by Fig.~\ref{SCSI_ICSI2}, Fig.~\ref{SCSI_ICSI3} shows that under a fixed $N/K$, the gains brought by ZFT/ZFR grow even though the average SNR ($\frac{P_{\rm{R}}}{2K}$) of each user decreases as the number of user pairs $K$ increases. These observations indicate that two asymptotic regimes, large-scale antenna arrays and high SNR, are equivalent~\cite{Ngo2013}. Moreover, larger gains over MRC/MRT are achieved by the ZFR/ZFT processing at higher $N/K$.

\begin{figure}
   \centering
   \includegraphics[scale=0.53]{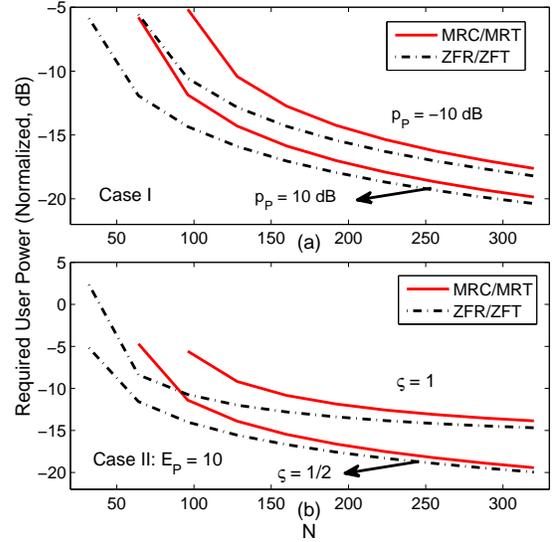}
   \caption{Required user power to achieve 1 bit/s/Hz per user for MRC/MRT and ZFR/ZFT (EPA, $P_{\rm R} =2K{P_{\rm S}}$).}
   \label{Asy_Ana1}
\end{figure}

Next the asymptotic analyses with massive arrays for the two cases in Propositions 1 and 2 are examined, supposing $P_{\rm R}=2K*P_{\rm S}$ and EPA employed with the transmit power at each user satisfying $p_i=P_{\rm S}$~($i=1,2,\cdots,2K$). Fig.~\ref{Asy_Ana1} shows the required user transmit power $P_{\rm S}$ to achieve $1$~bit/s/Hz per user. It is obvious from Fig.~\ref{Asy_Ana1}(a) that in case I where the pilot power $p_{\rm P}$ is fixed, the required user transmit power is significantly reduced as $N$ increases, and that the required $P_{\rm S}$ with ZFR/ZFT is lower than that with MRC/MRT. Regarding the imperfect CSI effect, less user transmit power is required when $p_{\rm P}$ is high. On the other hand, when $p_{\rm P}$ is low and $N$ is small, the required $1$~bit/s/Hz achievable rate per user cannot be achieved even with infinite $P_{\rm S}$, which means that the only way to reduce the imperfect CSI effect and thus achieve required spectral efficiency is to increase the number of antennas at the relay. For case II with $E_{\rm P}=10$ and the pilot power scaling down by ${p_{\rm P}} = \frac{{{E_{\rm{P}}}}}{N^\varsigma}$, Fig.~\ref{Asy_Ana1}(b) shows that higher $\varsigma$ leads to more slowly reduced $P_{\rm S}$, because the imperfect CSI effect becomes much severer when the pilot power is reduced faster with the increase of $N$.

\subsection{Power Allocation}
In this subsection, the proposed power allocation schemes in Section \ref{PA} are examined in regard to the performance of the spectral efficiency.

\begin{figure}
   \centering
   \includegraphics[scale=0.5]{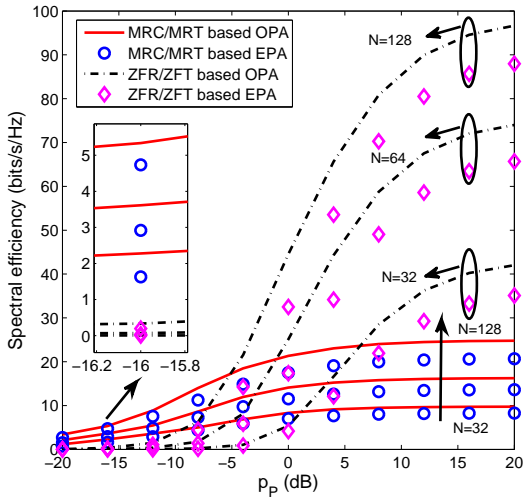}
   \caption{Spectral efficiency versus $p_{\rm P}$ ($P_0=10$~dB, $P_{\rm R,0}=23$~dB, $P=23$~dB).}
   \label{OPT3}
\end{figure}

In OPA, simulations are performed assuming that $P_0=10$~dB, $P_{\rm R,0}=23$~dB and $P=23$~dB. First, we choose the large-scale fading matrix as follows
\begin{equation*}
\begin{split}
{\bf D}=&diag\left\{ 0.749~0.045~0.246~0.121~0.125~0.142~0.635\right.\\
&~~~~~~\left.~0.256~0.021~0.123~0.257~0.856~1.000~0.899\right.\\
&~~~~~~\left.~0.014~0.759~0.315~0.432~0.195~0.562 \right\}
\end{split}
\end{equation*}
which is a snapshot of the practical setup, indicating that all large-scale fading factors fall into the interval $\left[ {0.014,1.000} \right]$. Fig.~\ref{OPT3} shows the spectral efficiency versus $p_{\rm P}$ with fixed $N=32$, $64$ and $128$ under both OPA and EPA. The employed EPA here allocates equal power to each user where the sum power consumed by all users achieves its maximum value $P/2$ with $P_{\rm R} = P/2$, i.e., $p_i=\frac{P}{{4K}}$ for all $i=1,2,\cdots,2K$. For the OPA, algorithm 1 is utilized with the initial values chosen as follows: $\varepsilon=0.01$, $L=10$, $\beta=1.1$, and $\chi_{k,1}={{{a_{k}}{P}} \over {\sum\limits_{i = 1}^{2K} {\left( {b_{k,i}^{(1)} + 2b_{k,i}^{(2)}P^{ - 1}} \right){P}}  + {c_{k}}{P} + \left( {4Kd_{k}^{(1)} + 8Kd_{k}^{(2)}P^{ - 1}} \right)}}$~($k=1,\cdots,2K$) are obtained by the EPA scheme. It can be observed from Fig.~\ref{OPT3} that OPA outperforms EPA, especially when the number of relay antennas is high, which demonstrates the effectiveness of our proposed PA. Furthermore, the spectral efficiency improvement in OPA for MRC/MRT beamforming is always smaller than that for ZFR/ZFT under different $p_{\rm P}$.

\begin{figure}
   \centering
   \includegraphics[scale=0.52]{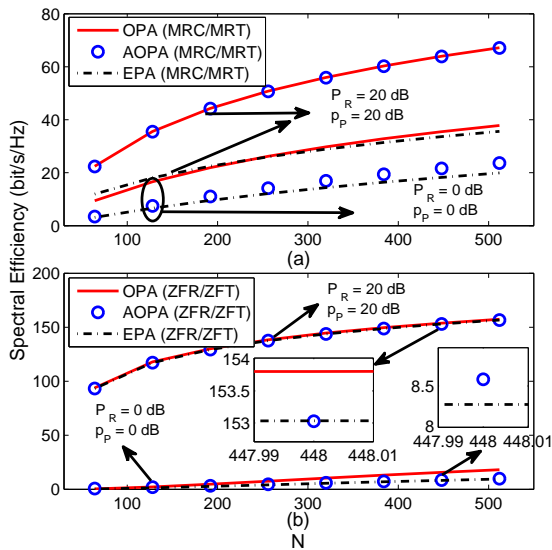}
   \caption{Spectral efficiency versus $N$ for different power allocation schemes (${P_{\rm R}}={P/2}$).}
   \label{DifOPA}
\end{figure}

For AOPA, Fig.~\ref{DifOPA} validates the effectiveness of (\ref{solution_app}) and (\ref{solution_app1}) at a large $N$ and a high SNR in comparison to OPA and EPA. Notably, the OPA here considers fixed ${P_{\rm R}}=P/2$, i.e., only user power allocation is performed. Fig.~\ref{DifOPA}(a) shows that when $P_{\rm R} = p_{\rm P}=20$~dB, the obtained spectral efficiency in AOPA for MRC/MRT\footnote{In the AOPA for MRC/MRT, we set $\sigma_{2i}^2=1/{\sigma_{2i-1}^2}$ for $i = 1,\cdots,K$ according to the assumption in Section~\ref{PA}.} is almost the same as that in OPA, reaching high gains over EPA. However, if the SNR is reduced to $P_{\rm R} = p_{\rm P}=0$~dB, the gain achieved by AOPA over EPA becomes rather slight even when the number of antennas is significantly increased. While regarding the ZFR/ZFT, Fig.~\ref{DifOPA}(b) illustrates that when the number of antennas is large and SNR is high, AOPA tends to EPA as predicted by (\ref{solution_app1}). Furthermore, Fig.~\ref{DifOPA}(b) shows that AOPA outperforms EPA slightly when the SNR is reduced to $P_{\rm R} = p_{\rm P}=0$~dB.

\section{Conclusion}\label{conclusion}
In this paper, closed-form ergodic achievable rate expressions have been derived for a multi-pair massive MIMO two-way AF relaying system with imperfect CSI and linear processing. Optimal power allocations schemes based on the obtained rate expressions have been shown to outperform equal power allocation in various scenarios. It has been found that the asymptotically optimal power solutions for MRC/MRT and ZFR/ZFT achieve almost the same performance as OPA when the SNR is high and the number of antennas at the relay is large. Both AOPA and OPA outperform EPA on the spectral efficiency. Besides, the allocated power of each user in AOPA is inverse to the large-scale fading factor of the channel from the user to the relay, and proportional to the channel from the relay to its destination for MRC/MRT under the condition that the link end-to-end large-scale fading factors among all pairs are equal, and the AOPA for ZFR/ZFT tends to be EPA when $N$ is large.

\section*{Appendix A: Proof of Equation (\ref{alpha})}\label{Appendix_A}
To prove (\ref{alpha}), we start from the expectation ${\rm E}\left[ {{{\left\| {{{\hat{\bf G}}^*}{\bf{T}}{{\hat{\bf G}}^H}{\bf{\tilde Gx}}} \right\|}^2}} \right]$, which is rewritten as (\ref{alpha_0}).
\begin{table*}
\begin{equation}\label{alpha_0}
\begin{split}
{\rm E}\left[ {{{\left\| {{{\hat{\bf G}}^*}{\bf{T}}{{\hat{\bf G}}^H}{\tilde{\bf G}}{\bf x}} \right\|}^2}} \right]\mathop  = \limits^{(a)}~&{\rm Tr}\left\{ {{\rm E}\left[ {{{\hat{\bf G}}^*}{\bf{T}}{{\hat{\bf G}}^H}\left( {{\hat {\bf G}} - {\bf{\Xi}}} \right){\bf{P}}{{\bf{P}}^H}{{\left( {{\hat{\bf G}} - {\bf{\Xi }}} \right)}^H}{\hat {\bf G}}{{\bf{T}}^H}{{\hat{\bf G}}^T}} \right]} \right\}\\
\mathop  = \limits^{(b)}~&{\rm Tr}\left\{ {{\rm E}\left[ {{\hat {\bf G}}{\bf{P}}{{\bf{P}}^H}{{\hat{\bf G}}^H}}{{\hat{\bf G}}{{\bf{T}}^H}{{\hat{\bf G}}^T}{{\hat{\bf G}}^*}{\bf{T}}{{\hat{\bf G}}^H}} \right]} \right\}+{\rm Tr}\left\{ {{\rm E}\left[{{\bf{\Xi }}{\bf{P}}{{\bf{P}}^H}{{\bf{\Xi }}^H}} {{\hat{\bf G}}{{\bf{T}}^H}{{\hat{\bf G}}^T}{{\hat{\bf G}}^*}{\bf{T}}{{\hat{\bf G}}^H}} \right]} \right\}
\end{split}
\end{equation}
\begin{equation}\label{alpha_1_main}
\begin{split}
{\rm Tr}\left\{ {{\rm E}\left[ {{\hat {\bf G}}{\bf{P}}{{\bf{P}}^H}{{\hat{\bf G}}^H}}{\hat{\bf G}}{{\bf{T}}^H}{{\hat{\bf G}}^T}{{\hat{\bf G}}^*}{\bf{T}}{{\hat{\bf G}}^H} \right]} \right\}\mathop  = \limits^{(a)}&{\rm Tr}\left\{ {\sum\limits_{k = 1}^{2K}{p_{k}} {{\rm E}\left[{{\hat{\bf g}}_{k}}{\hat{\bf g}}_{k}^H{\sum\limits_{i = 1}^K {\left( {{{\hat{\bf g}}_{2i - 1}}{\hat{\bf g}}_{2i}^T + {{\hat{\bf g}}_{2i}}{\hat{\bf g}}_{2i - 1}^T} \right)\left( {{\hat{\bf g}}_{2i - 1}^*{\hat{\bf g}}_{2i}^H + {\hat{\bf g}}_{2i}^*{\hat{\bf g}}_{2i - 1}^H} \right)} } \right]} }\right\}\\
\mathop  = \limits^{(b)}& {2N\left( {N + 1} \right)\hat\Psi\hat\Phi  + N{{\left( {N + 1} \right)}^2}\sum\limits_{i = 1}^K{\hat\psi _i}}\hat\phi_i
\end{split}
\end{equation}
\begin{equation}\label{alpha_1_ext1}
\begin{split}
{\rm Tr}\left\{ {{\rm E}\left[ {{\hat{\bf g}}_k}{\hat{\bf g}}_k^H{{{\hat{\bf g}}_{2i}}{\hat{\bf g}}_{2i - 1}^T{\hat{\bf g}}_{2i - 1}^*{\hat{\bf g}}_{2i}^H} \right]} \right\}
=~{\rm Tr}\left\{ {\rm E}\left[ {{{\hat{\bf g}}_k}{\hat{\bf g}}_k^H} \right]{\rm E}\left[ {{{\hat{\bf g}}_{2i}}{\hat{\bf g}}_{2i - 1}^H}{{{\hat{\bf g}}_{2i-1}}{\hat{\bf g}}_{2i}^H} \right] \right\} = {N^2}\hat\sigma _{2i - 1}^2\hat\sigma _{2i}^2\hat\sigma _k^2
\end{split}
\end{equation}
\begin{equation}\label{alpha_1_ext2}
\begin{split}
{\rm Tr}\left\{ {{\rm E}\left[ {{\hat{\bf g}}_k}{\hat{\bf g}}_k^H{{{\hat{\bf g}}_{2i - 1}}{\hat{\bf g}}_{2i}^T{\hat{\bf g}}_{2i - 1}^*{\hat{\bf g}}_{2i}^H} \right]} \right\}
=~{\rm Tr}\left\{ {\rm E}\left[ {{{\hat{\bf g}}_k}{\hat{\bf g}}_k^H} \right]{\rm E}\left[ {{{\hat{\bf g}}_{2i - 1}}{\hat{\bf g}}_{2i - 1}^H} \right]{\rm E}\left[ {{{\hat{\bf g}}_{2i}}{\hat{\bf g}}_{2i}^H} \right] \right\} = N\hat\sigma _{2i - 1}^2\hat\sigma _{2i}^2\hat\sigma _k^2
\end{split}
\end{equation}
\begin{equation}\label{alpha_1_ext3}
\begin{split}
{\rm Tr}\left\{ {{\rm E}\left[ {{\hat{\bf g}}_{2i - 1}{\hat{\bf g}}_{2i-1}^H\left({{\hat{\bf g}}_{2i - 1}}{\hat{\bf g}}_{2i}^T+{{\hat{\bf g}}_{2i}}{\hat{\bf g}}_{2i-1}^T\right)\left( {{{\hat{\bf g}}_{2i - 1}^*}{\hat{\bf g}}_{2i}^H + {{\hat{\bf g}}_{2i}^*}{\hat{\bf g}}_{2i-1}^H} \right)} \right]} \right\}=~2N\left( {N + 1} \right)\hat \sigma _{2i}^2\hat \sigma _{2i - 1}^4 + {N}\left( {N + 1} \right)^2\hat \sigma _{2i}^2\hat \sigma _{2i - 1}^4
\end{split}
\end{equation}
\begin{equation}\label{alpha_1_ext4}
\begin{split}
{\rm Tr}\left\{ {{\rm E}\left[ {{\hat{\bf g}}_{2i}{\hat{\bf g}}_{2i}^H\left({{\hat{\bf g}}_{2i - 1}}{\hat{\bf g}}_{2i}^T+{{\hat{\bf g}}_{2i}}{\hat{\bf g}}_{2i-1}^T\right)\left( {{{\hat{\bf g}}_{2i - 1}^*}{\hat{\bf g}}_{2i}^H + {{\hat{\bf g}}_{2i}^*}{\hat{\bf g}}_{2i-1}^H} \right)} \right]} \right\}=~ 2N\left( {N + 1} \right)\hat \sigma _{2i - 1}^2\hat \sigma _{2i}^4 + {N}\left( {N + 1} \right)^2\hat \sigma _{2i - 1}^2\hat \sigma _{2i}^4
\end{split}
\end{equation}
\begin{equation}\label{alpha_1_error}
\begin{split}
{\rm Tr}\left\{ {{\rm E}\left[{{\bf{\Xi }}{\bf{P}}{{\bf{P}}^H}{{\bf{\Xi }}^H}} {{\hat{\bf G}}{{\bf{T}}^H}{{\hat{\bf G}}^T}{{\hat{\bf G}}^*}{\bf{T}}{{\hat{\bf G}}^H}} \right]} \right\}= ~ &\sum\limits_{j = 1}^{2K} {{p_j}\sigma _{{\xi _j}}^2}{\rm Tr}\left\{{{\rm E}\left[{\sum\limits_{i = 1}^K {\left( {{{\hat{\bf g}}_{2i - 1}}{\hat{\bf g}}_{2i}^T + {{\hat{\bf g}}_{2i}}{\hat{\bf g}}_{2i - 1}^T} \right)\left( {{\hat{\bf g}}_{2i - 1}^*{\hat{\bf g}}_{2i}^H + {\hat{\bf g}}_{2i}^*{\hat{\bf g}}_{2i - 1}^H} \right)} } \right]}\right\}\\
= ~ &2N\left( {N + 1} \right)\hat\Phi\sum\limits_{j = 1}^{2K} {{p_j}\sigma _{{\xi _j}}^2}.
\end{split}
\end{equation}
\end{table*}
Step $(a)$ is obtained by ${\rm E}\left[ {{\bf{x}}{{\bf{x}}^H}} \right] = {{\bf{I}}_{2K}}$ and substituting $\tilde {\bf G} = {\bf GP}$ and (\ref{estimated_CSI_exp}) into the equation, and step $(b)$ results from the independence between $\hat{\bf G}$ and $\bf\Xi$ and the property ${\rm Tr}\left\{ {{\bf{AB}}} \right\} = {\rm Tr}\left\{ {{\bf{BA}}} \right\}$. As for the first term in (\ref{alpha_0}), we have (\ref{alpha_1_main}), where $\hat\Psi  = \sum\limits_{i = 1}^K {{\hat\psi _i}}$ with ${\hat\psi _i} = {p_{2i - 1}}\hat\sigma _{2i - 1}^2 + {p_{2i}}\hat\sigma _{2i}^2$ and $\hat \Phi  = \sum\limits_{i = 1}^K {{\hat \phi _i}}$ with ${\hat \phi _i} = \hat \sigma _{2i - 1}^2\hat \sigma _{2i}^2$. Step $(a)$ in (\ref{alpha_1_main}) results from substituting $\hat{\bf G}$ in (\ref{estimated_CSI_exp}), ${{\bf P}}$ and ${\bf T}$ into the equation and formula expansion based on the fact that the expectation of $\sum\limits_{k=1}^{2K}{{\rm E}\left[ {\hat{\bf g}}_{k}{\hat{\bf g}}_{k}^H{{\left( {{{\hat{\bf g}}_{2j - 1}}{\hat{\bf g}}_{2j}^T + {{\hat{\bf g}}_{2j}}{\hat{\bf g}}_{2j - 1}^T} \right)}\left( {{\hat{\bf g}}_{2i - 1}^*{\hat{\bf g}}_{2i}^H + {\hat{\bf g}}_{2i}^*{\hat{\bf g}}_{2i - 1}^H} \right) } \right]}$ $=0$ for any $i\ne j$. Step $(b)$ is obtained by ${\rm Tr}\left\{ {{\bf{AB}}} \right\} = {\rm Tr}\left\{ {{\bf{BA}}} \right\}$, some results from Gaussian distributed estimated channel in (\ref{estimated_CSI_exp})\footnote{Due to the estimated channel model in (\ref{estimated_CSI_exp}), we have that $\hat{\bf g}_i$ and $\hat{\bf g}_j$ are mutually independent $N\times 1$ vectors with $\forall i\ne j$ whose elements are i.i.d. zero-mean Gaussian distributed with variances $\hat\sigma_i^2$ and $\hat\sigma_j^2$, respectively. Then, it can be concluded that ${\rm E}\left[{\hat{\bf{g}}_i^H}\hat{\bf{g}}_i\right]=N\hat\sigma _i^2$, ${\rm E}\{{\hat{\bf{g}}_j^H}\hat{\bf{g}}_j\}=N\hat\sigma _j^2$, and ${\rm E}\left[{\hat{\bf{g}}_i^H}\hat{\bf{g}}_j\right]=0$. Also, we can obtain that ${\rm E}\left[|{\hat{\bf{g}}_i^H}\hat{\bf{g}}_j|^2\right]=N\hat\sigma _i^2\hat\sigma _j^2$.}~\cite{Cramer1970}, and \cite[Lemma~2.9]{Tulino2004}. To elaborate in detail, for the items in $(a)$ with any $k\ne 2i-1$ or $2i$, we have (\ref{alpha_1_ext1}) and (\ref{alpha_1_ext2}), where the properties ${{\bf x}^*{\bf y}^T}={{\bf y}{\bf x}^H}$ and ${{\bf x}^T{\bf y}^*}={{\bf y}^H{\bf x}}$ for arbitrary vectors ${\bf x}$ and ${\bf y}$ are utilized. While for the items with $k= 2i-1$ and $2i$, we have (\ref{alpha_1_ext3}) and (\ref{alpha_1_ext4}). On account of the independence between $\hat {\bf G}$ and $\bf \Xi$, the second term in (\ref{alpha_0}) becomes (\ref{alpha_1_error}). By substituting (\ref{alpha_1_main}) and (\ref{alpha_1_error}) into (\ref{alpha_0}), we have
\begin{equation}\label{alpha_1}
\begin{split}
&{\rm E}\left[ {{\left\| {{{\hat{\bf G}}^*}{\bf{T}}{{\hat{\bf G}}^H}{\tilde{\bf G}}{\bf x}} \right\|}^2} \right]=N\left( {N + 1} \right)\left[2\Psi\hat\Phi  + \left( {N + 1} \right)\sum\limits_{i = 1}^K{\hat\psi _i}\hat\phi_i\right]
\end{split}
\end{equation}
where $\Psi  = \sum\limits_{i = 1}^K {\psi_i}$ with $\psi_i={{{p_{2i - 1}}{\sigma} _{2i - 1}^2 + {p_{2i}}{\sigma} _{2i}^2}}$.

To proceed, we need to calculate the expectation ${\rm E}\left[ {{{\left\| {{{\hat{\bf G}}^*}{\bf{T}}{{\hat{\bf G}}^H}{{\bf{n}}_{\rm{r}}}} \right\|}^2}} \right]$ in (\ref{alpha}), which is elaborated as follows
\begin{equation}\label{alpha_2}
\begin{split}
& {\rm E}\left[ {{{\left\| {{{\hat{\bf G}}^*}{\bf{T}}{{\hat{\bf G}}^H}{{\bf{n}}_{\rm{r}}}} \right\|}^2}} \right]\mathop  = \limits^{(a)}{\rm Tr}\left\{ {\sigma _n^2{\rm E}\left[ {{{\hat{\bf G}}^*}{\bf{T}}{{\hat{\bf G}}^H}{\hat{\bf G}}{{\bf{T}}^H}{{\hat{\bf G}}^T}} \right]} \right\} \mathop  = \limits^{(b)}\\
&{\rm Tr}\left\{ \sigma _n^2{{\rm E}\left[ {\sum\limits_{i = 1}^K {\left( {{\hat{\bf g}}_{2i - 1}^*{\hat{\bf g}}_{2i}^H + {\hat{\bf g}}_{2i}^*{\hat{\bf g}}_{2i - 1}^H} \right)\left( {{{\hat{\bf g}}_{2i - 1}}{\hat{\bf g}}_{2i}^T + {{\hat{\bf g}}_{2i}}{\hat{\bf g}}_{2i - 1}^T} \right)} } \right]} \right\}\\
&\mathop  = \limits^{(c)} 2N\left( {N + 1} \right)\sigma _n^2\hat\Phi
\end{split}
\end{equation}
where step $(a)$ is obtained by ${\rm E}\left[ {{{\bf{n}}_{\rm{r}}}{\bf{n}}_{\rm{r}}^H} \right] = \sigma _n^2{{\bf{I}}_N}$ and ${\rm Tr}\left\{ {{\bf{AB}}} \right\} = {\rm Tr}\left\{ {{\bf{BA}}} \right\}$, $(b)$ results from substituting $\hat{\bf G}$ in (\ref{estimated_CSI_exp}), $\bf P$ and ${\bf T}$ into the equation and the fact that the expectation of ${{\rm E}\left[ {\left( {{\hat{\bf g}}_{2i - 1}^*{\hat{\bf g}}_{2i}^H + {\hat{\bf g}}_{2i}^*{\hat{\bf g}}_{2i - 1}^H} \right){\left( {{{\hat{\bf g}}_{2j - 1}}{\hat{\bf g}}_{2j}^T + {{\hat{\bf g}}_{2j}}{\hat{\bf g}}_{2j - 1}^T} \right)} } \right]}=0$ for any $i\ne j$, and step $(c)$ results from the properties of ${\rm Tr}\left\{ {{\bf{AB}}} \right\} = {\rm Tr}\left\{ {{\bf{BA}}} \right\}$, ${{\bf x}^*{\bf y}^T}={{\bf y}{\bf x}^H}$ and ${{\bf x}^T{\bf y}^*}={{\bf y}^H{\bf x}}$ for arbitrary vectors ${\bf x}$ and ${\bf y}$, and the properties of Gaussian distributed vectors, respectively, the detailed derivation of which is
\begin{equation}
\begin{split}
& {\rm Tr}\left\{ {\rm E}\left[ {\left( {{\hat{\bf g}}_{2i - 1}^*{\hat{\bf g}}_{2i}^H + {\hat{\bf g}}_{2i}^*{\hat{\bf g}}_{2i - 1}^H} \right)\left( {{{\hat{\bf g}}_{2i - 1}}{\hat{\bf g}}_{2i}^T + {{\hat{\bf g}}_{2i}}{\hat{\bf g}}_{2i - 1}^T} \right)} \right]\right\}\\
=~& 2{\rm E}\left[ {{\hat{\bf g}}_{2i}^H{{\hat{\bf g}}_{2i - 1}}{\hat{\bf g}}_{2i - 1}^H{{\hat{\bf g}}_{2i}}} \right] + 2{\rm E}\left[ {{\hat{\bf g}}_{2i}^H{{\hat{\bf g}}_{2i}}} \right]{\rm E}\left[ {{\hat{\bf g}}_{2i - 1}^H{{\hat{\bf g}}_{2i - 1}}} \right]\\
=~& 2N\left( {N + 1} \right)\hat\sigma _{2i - 1}^2\hat\sigma _{2i}^2.
\end{split}
\end{equation}
Hence, by substituting (\ref{alpha_1}) and (\ref{alpha_2}) into the step $(a)$ in (\ref{alpha}), the proof of (\ref{alpha}) is completed.

\section*{Appendix B: Proof of Equation (\ref{hat_alpha})}\label{Appendix_B}
To prove (\ref{hat_alpha}), likewise, we start from ${\rm E}\left[ {{{\left\| {{{\hat{ \bar{\bf G}}}^*}{\bf{T}}{{\hat {\bar {\bf G}}}^H}{\tilde{\bf G}}{\bf x}} \right\|}^2}} \right]$, which can be rewritten as (\ref{hat_alpha_1}),
\begin{table*}
\begin{equation}\label{hat_alpha_1}
\begin{split}
&{\rm E}\left[ {{{\left\| {{{\hat{ \bar{\bf G}}}^*}{\bf{T}}{{\hat {\bar {\bf G}}}^H}{\tilde{\bf G}}{\bf x}} \right\|}^2}} \right] = {\rm Tr}\left\{ {{\rm E}\left[ {{{\hat{ \bar{\bf G}}}^*}{\bf{T}}{{\hat{ \bar{\bf G}}}^H}\left( {{\hat{\bf G}} - {\bf{\Xi }}} \right){\bf{P}}{{\bf{P}}^H}{{\left( {{\hat{\bf G}} - {\bf{\Xi }}} \right)}^H}{\hat{ \bar{\bf G}}}{\bf{T}}{{\hat{ \bar{\bf G}}}^T}} \right]} \right\}\mathop  = \limits^{(a)}{\rm Tr}\left\{ {{\rm E}\left[ {{{\hat{ \bar{\bf G}}}^*}{\bf{T}}{{\hat{ \bar{\bf G}}}^H}{\hat{{\bf G}}}{\bf{P}}{{\bf{P}}^H}{{{\hat{\bf G}}}^H}{\hat{ \bar{\bf G}}}{\bf{T}}{{\hat{ \bar{\bf G}}}^T}} \right]} \right\}\\
&~~~+ {\rm Tr}\left\{ {{\rm E}\left[ {{\bf{\Xi P}}{{\bf{P}}^H}{{\bf{\Xi }}^H}{\hat{ \bar{\bf G}}}{\bf{T}}{{\hat{ \bar{\bf G}}}^T}{{\hat{ \bar{\bf G}}}^*}{\bf{T}}{{\hat{ \bar{\bf G}}}^H}} \right]} \right\}\mathop  = \limits^{(b)}~ {\rm Tr}\left\{ {{\rm E}\left[ {{{\hat{ \bar{\bf G}}}^*}{{\bf{P}}_T}{{\hat{ \bar{\bf G}}}^T}} \right]} \right\} + \sum\limits_{i = 1}^{2K} {{p_i}\sigma _{{\xi _i}}^2} {\rm Tr}\left\{ {{\rm E}\left[ {{{{\hat{\bf \Omega}}}^*}{{\bf T}{\hat {\bf \Omega}}{\bf T}}} \right]} \right\}\\
&\mathop  = \limits^{(c)}~ \sum\limits_{i = 1}^{2K} {{p_{i'}}} {\rm E}\left[ {{\hat{ \bar{\bf g}}}_i^H{{\hat{ \bar{\bf g}}}_i}} \right] + \sum\limits_{i = 1}^{2K} {{p_i}\sigma _{{\xi _i}}^2} \sum\limits_{j = 1}^{2K} {\left( {{\rm E}\left[ {\hat \omega_{j,j}^*{{\hat \omega}_{j',j'}}} \right] + {\rm E}\left[ {{{\left| {{{\hat \omega }_{j,j'}}} \right|}^2}} \right]} \right)} \mathop  = \limits^{(d)}~ \sum\limits_{i = 1}^{2K} {\frac{{{p_{i'}}}}{{\left( {N - 2K - 1} \right)\hat \sigma _i^2}}}  + \hat \eta {\sum\limits_{i = 1}^{2K} {{p_i}\sigma _{{\xi _i}}^2} }
\end{split}
\end{equation}
\end{table*}
where $\hat \eta  = \sum\limits_{j = 1}^{2K} {\frac{1}{{\left( {N - 2K} \right)\left( {N - 2K - 3} \right)\hat \sigma _j^2\hat \sigma _{j'}^2}}}$, step $(a)$ is based on the fact that the estimation error matrix ${\bf{\Xi }}$ is independent of ${{\bf{\hat G}}}$, and step $(b)$ is obtained by an intuitive property of ${{\hat{\bar{\bf G}}}^H}{\hat{\bf{G}}} = {\hat{\bf{G}}^H}{\hat{\bar{\bf G}}} = {{\bf{I}}}_{2K}$, the definition of ${{\bf{P}}_T} = {\bf{TP}}{{\bf{P}}^H}{\bf{T}} = diag\left\{ {{p_2},{p_1}, \ldots {p_{2K}},{p_{2K - 1}}} \right\}$ and ${{\rm E}\left[ {{\bf{\Xi P}}{{\bf{P}}^H}{{\bf{\Xi }}^H}} \right]}=\sum\limits_{i = 1}^{2K} {{p_i}\sigma _{{\xi _i}}^2}{{\bf{I}}}_{N}$, which is derived from the distribution of $\bf{\Xi}$. Here, ${\hat {\bf \Omega }} \buildrel \Delta \over = {{\left( {{\hat {\bf{G}}^H}\hat {\bf{G}}} \right)}^{ - 1}}$ is defined with $\hat {\omega }_{i,j} = {\left( \hat {\bf{\Omega }} \right)_{i,j}}$ for $\forall i,j\in\left\{1,2,\cdots,2K\right\}$, following an inverse Wishart distribution of $\mathcal{W}_{2K}^{-1}(N+2K+1,\hat {\bf{D }}^{-1})$. Furthermore, step $(c)$ is based on ${\rm Tr}\left\{ {{\bf{AB}}} \right\} = {\rm Tr}\left\{ {{\bf{BA}}} \right\}$. As to the detailed derivation of $(d)$, we use the identity~\cite{Timm2002,Graczyk2003}
\begin{equation}\label{inv_Wishart_definition}
{\rm E}\left[  {\bf{W}}^{ - 1} \right] = \frac{{{{\bf{\Sigma }}^{ - 1}}}}{{n - m - 1}}
\end{equation}
where ${\bf{W}} \sim \mathcal{W}_m(n,{\bf{\Sigma }})$ is an $m\times m$ central complex Wishart matrix with $n$~($n> m$) degrees of freedom and the distribution of ${\bf{W}}^{ - 1}$ is called an inverted Wishart distribution, following $\mathcal{W}_m^{-1}(n+m+1,{\bf{\Sigma }}^{-1})$. It can be easily concluded that $\hat{\bf{\Omega }} \buildrel \Delta \over = {{\left( {{{\hat{\bf G}}^H}\hat{\bf{G}}} \right)}^{ - 1}}\sim \mathcal{W}_{2K}^{-1}(N+2K+1,\hat{\bf{D }}^{-1})$ with ${\hat\omega _{i,j}} = {\left( \hat{\bf{\Omega }} \right)_{i,j}}$ for $\forall i,j\in\left\{1,2,\cdots,2K\right\}$, hence
\begin{equation}\label{inv_Wishart}
{\rm E}\left[ {{{\hat{\bar{\bf G}}}^H}\hat{\bar{\bf G}}} \right] = {\rm E}\left[ {{{\left( {{\hat{\bf{G}}^H}\hat{\bf{G}}} \right)}^{ - 1}}} \right] = \frac{{{\hat{\bf{D}}^{ - 1}}}}{{N - 2K - 1}}
\end{equation}
where $N>2K$. In this way, we have ${\rm E}\left[ {\hat{\bar{\bf g}}_k^H{{\hat{\bar{\bf g}}}_k}} \right]={\rm E}\left[{\hat\omega _{k,k}}\right]={\frac{1}{{\left( {N - 2K - 1} \right)\hat\sigma _k^2}}}$ for $k=1,2,\cdots,2K$, (\ref{hat_alpha2_ext1}) and (\ref{hat_alpha2_ext2}), where ${\mathop{\rm cov}} \left[ {\hat \omega_{k,k}^*{\hat \omega_{k',k'}}} \right]$~($k,k'\in\left\{1,2,\cdots,2K\right\}$), ${\rm E}\left[ {\hat \omega_{i,j}} \right]$ and ${\mathop{\rm var}} \left[ {{\hat \omega_{i,j}}} \right]$~($i,j\in\left\{1,2,\cdots,2K\right\}$) are calculated based on the properties of the inverse Wishart matrix ${\bf {\hat \Omega}}$~\cite{Timm2002,Graczyk2003}.

Then, the calculation of ${\rm E}\left[ {{{\left\| {{{\hat{ \bar{\bf G}}}^*}{\bf{T}}{{\hat{ \bar{\bf G}}}^H}{{\bf{n}}_{\rm{r}}}} \right\|}^2}} \right]$ in (\ref{hat_alpha}) can be elaborated as (\ref{hat_alpha_2}). Hence, by substituting (\ref{hat_alpha_1}) and (\ref{hat_alpha_2}) into the step $(a)$ in (\ref{hat_alpha}), the proof of (\ref{hat_alpha}) is completed.

\section*{Appendix C: Proof of Theorem 1}\label{Appendix_C}
To derive the closed-form expression of the achievable rate in (\ref{SCSI_rate}), we start from the expectation ${\rm E}\left[ {{\bf{g}}_{k'}^T{\bf{F}}{{\bf{g}}_k}} \right]$ in the numerator based on MRC/MRT in (\ref{F}), given by (\ref{mean_upp}), where step $(a)$ is obtained by ${{\bf g}}_{k'}={\hat{\bf g}}_{k'}-{\bf \xi}_{k'}$ and the independence between ${\hat{\bf g}}_{k'}$ and ${\bf \xi}_{k'}$, step $(b)$ results from substituting ${\bf F}$ in (\ref{F}), ${\hat{\bf G}}$ in (\ref{estimated_CSI_exp}) and ${\bf T}$ into the equation and the fact ${\rm E}\left[ {{\bf{\xi }}_{k'}^T{\bf{F}}{{\bf{\xi }}_k}} \right]=0$, $(c)$ is obtained by formula expansion based on the fact that the expectation of ${\rm E}\left[ {{\hat{\bf g}}_{k'}^T\left( {{\hat{\bf g}}_{2i - 1}^*{\hat{\bf g}}_{2i}^H + {\hat{\bf g}}_{2i}^*{\hat{\bf g}}_{2i - 1}^H} \right){{\hat{\bf g}}_k}} \right]=0$ for any $i \ne \left\lceil {\frac{k}{2}} \right\rceil$ and $\left\lceil {\frac{{k'}}{2}} \right\rceil$, and step $(d)$ results from the property ${\rm Tr}\left\{ {{\bf{AB}}} \right\} = {\rm Tr}\left\{ {{\bf{BA}}} \right\}$ and the properties of Gaussian distributed vectors.
\begin{table*}
\begin{equation}\label{hat_alpha2_ext1}
\begin{split}
&{\rm E}\left[ {\hat \omega _{j,j}^*{{\hat \omega }_{j',j'}}} \right] = {\rm{cov}}\left[ {\hat \omega _{j,j}^*{{\hat \omega }_{j',j'}}} \right] + {\rm E}\left[ {\hat \omega _{j,j}^*} \right]{\rm E}\left[ {{{\hat \omega }_{j',j'}}} \right]\\
=~&\frac{2}{{\left( {N - 2K} \right){{\left( {N - 2K - 1} \right)}^2}\left( {N - 2K - 3} \right){\hat \sigma} _j^2{\hat \sigma} _{j'}^2}}+\frac{1}{{{{\left( {N - 2K - 1} \right)}^2}{\hat \sigma} _j^2{\hat \sigma} _{j'}^2}}
\end{split}
\end{equation}
\begin{equation}\label{hat_alpha2_ext2}
\begin{split}
{\rm E}\left[ {{{\left| {{\hat \omega_{i,j}}} \right|}^2}} \right] =~{\mathop{\rm var}} \left[ {{\hat \omega_{i,j}}} \right] + {\rm E}^2\left[ {{\hat \omega_{i,j}}} \right] =~ \frac{2}{{{{\left( {N - 2K - 1} \right)}^2}\left( {N - 2K - 3} \right)\hat \sigma _i^2\hat \sigma _j^2}}
\end{split}
\end{equation}
\begin{equation}\label{hat_alpha_2}
\begin{split}
{\rm E}\left[ {{\left\| {{{\hat{ \bar{\bf G}}}^*}{\bf{T}}{{\hat{ \bar{\bf G}}}^H}{{\bf{n}}_{\rm{r}}}} \right\|}^2} \right] = ~{\rm Tr}\left\{ {\sigma _n^2{\rm E}\left[ {{{{{\hat{\bf \Omega}}}^*}{{\bf T}{\hat {\bf \Omega}}{\bf T}}}} \right]} \right\}= ~ \sigma _n^2\sum\limits_{j = 1}^{2K} {\left( {{\rm E}\left[ {\hat \omega_{j,j}^*{{\hat \omega}_{j',j'}}} \right] + {\rm E}\left[ {{{\left| {{\hat \omega_{i,j}}} \right|}^2}} \right]} \right)} = ~\hat \eta {\sigma _n^2}.
\end{split}
\end{equation}
\begin{equation}\label{mean_upp}
\begin{split}
{\rm E}\left[ {{\bf{g}}_{k'}^T{\bf{F}}{{\bf{g}}_k}} \right] \mathop  = \limits^{(a)}~& {\rm E}\left[ {{\hat{\bf g}}_{k'}^T{\bf{F}}{{\hat{\bf g}}_k}} \right] + {\rm E}\left[ {{\bf{\xi }}_{k'}^T{\bf{F}}{{\bf{\xi }}_k}} \right]\mathop = \limits^{(b)} \alpha_1 {\rm E}\left[ {{\hat{\bf g}}_{k'}^T\sum\limits_{i = 1}^K {\left( {{\hat{\bf g}}_{2i - 1}^*{\hat{\bf g}}_{2i}^H + {\hat{\bf g}}_{2i}^*{\hat{\bf g}}_{2i - 1}^H} \right)} {{\hat{\bf g}}_k}} \right]\\
\mathop  = \limits^{(c)}~& \alpha_1 {\rm E}\left[ {{\hat{\bf g}}_{k'}^T{\hat{\bf g}}_k^*{\hat{\bf g}}_{k'}^H{{\hat{\bf g}}_k} + {\hat{\bf g}}_{k'}^T{\hat{\bf g}}_{k'}^*{\hat{\bf g}}_k^H{{\hat{\bf g}}_k}} \right] \mathop  = \limits^{(d)} \alpha_1 N\left( {N + 1} \right)\hat\phi_{\left\lceil {\frac{{k'}}{2}} \right\rceil}
\end{split}
\end{equation}
\begin{equation}\label{var_down_main}
\begin{split}
{\rm Var}\left[ {{{\bf g}}_{k'}^T{\bf{F}}{{{\bf g}}_k}} \right] \mathop  = \limits^{(a)}& {\rm E}\left[ {{{\left| {{{\bf g}}_{k'}^T{\bf{F}}{{{\bf g}}_k}} \right|}^2}} \right] - {\left| {{\rm E}\left[ {{{\bf g}}_{k'}^T{\bf{F}}{{{\bf g}}_k}} \right]} \right|^2}\mathop  = \limits^{(b)} {{\rm E}\left[ {{\hat{\bf g}}_{k'}^T{\bf F}{{\hat{\bf g}}_k}{\hat{\bf g}}_k^H{\bf F}^H{\hat{\bf g}}_{k'}^*} \right]}+{{\rm E}\left[ {{\hat{\bf g}}_{k'}^T{\bf F}{{{\bf \xi}}_k}{{\bf \xi}}_k^H{\bf F}^H{\hat{\bf g}}_{k'}^*} \right]}\\
&+{{\rm E}\left[ {{{\bf \xi}}_{k'}^T{\bf F}{{\hat{\bf g}}_k}{\hat{\bf g}}_k^H{\bf F}^H{{\bf \xi}}_{k'}^*} \right]}+{{\rm E}\left[ {{{\bf \xi}}_{k'}^T{\bf F}{{{\bf \xi}}_k}{{\bf \xi}}_k^H{\bf F}^H{{\bf \xi}}_{k'}^*} \right]} - {\alpha_1 ^2}{N^2}{\left( {N + 1} \right)^2}\hat\phi_{\left\lceil {\frac{{k'}}{2}} \right\rceil}^2
\end{split}
\end{equation}
\begin{equation}\label{var_down_main1}
\begin{split}
{{\rm E}\left[ {{\hat{\bf g}}_{k'}^T{\bf F}{{\hat{\bf g}}_k}{\hat{\bf g}}_k^H{\bf F}^H{\hat{\bf g}}_{k'}^*} \right]}\mathop  = \limits^{(a)}~&{\alpha_1 ^2}\sum\limits_{i = 1}^K {{\rm E}\left[ {{\hat{\bf g}}_{k'}^T\left( {{\hat{\bf g}}_{2i - 1}^*{\hat{\bf g}}_{2i}^H + {\hat{\bf g}}_{2i}^*{\hat{\bf g}}_{2i - 1}^H} \right){{\hat{\bf g}}_k}{\hat{\bf g}}_k^H\left( {{{\hat{\bf g}}_{2i - 1}}{\hat{\bf g}}_{2i}^T + {{\hat{\bf g}}_{2i}}{\hat{\bf g}}_{2i - 1}^T} \right){\hat{\bf g}}_{k'}^*} \right]}\\
\mathop  = \limits^{(b)}~ & 2N\left( {N+1} \right)^2{\alpha_1 ^2}\hat\phi_{\left\lceil {\frac{{k'}}{2}} \right\rceil}^2+2N\left( {N+1} \right){\alpha_1 ^2}\hat\phi_{\left\lceil {\frac{{k'}}{2}} \right\rceil}\hat\Phi+N^2\left(N+1\right)^2{\alpha_1 ^2}\hat\phi_{\left\lceil {\frac{{k'}}{2}} \right\rceil}^2
\end{split}
\end{equation}
\begin{equation}\label{var_down_ext1}
\begin{split}
&{\rm E}\left[ {{\hat{\bf g}}_{k'}^T\left( {{\hat{\bf g}}_k^*{\hat{\bf g}}_{k'}^H + {\hat{\bf g}}_{k'}^*{\hat{\bf g}}_k^H} \right){{\hat{\bf g}}_k}{\hat{\bf g}}_k^H\left( {{{\hat{\bf g}}_k}{\hat{\bf g}}_{k'}^T + {{\hat{\bf g}}_{k'}}{\hat{\bf g}}_k^T} \right){\hat{\bf g}}_{k'}^*} \right]=~ 2{\rm Tr}\left\{ {{\rm E}\left[ {{{\hat{\bf g}}_{k'}}{\hat{\bf g}}_{k'}^H{{\hat{\bf g}}_{k'}}{\hat{\bf g}}_{k'}^H} \right]{\rm E}\left[ {{{\hat{\bf g}}_k}{{\hat{\bf g}}_k}^H{{\hat{\bf g}}_k}{\hat{\bf g}}_k^H} \right]} \right\}+ {\rm E}\left[ {{{\left| {{\hat{\bf g}}_k^H{{\hat{\bf g}}_{k'}}} \right|}^4}} \right] \\
 &~~+ {\rm E}\left[ {{{\left\| {{{\hat{\bf g}}_k}} \right\|}^4}} \right]{\rm E}\left[ {{{\left\| {{{\hat{\bf g}}_{k'}}} \right\|}^4}} \right]=~2N\left( {N+1} \right)^2\hat\phi_{\left\lceil {\frac{{k'}}{2}} \right\rceil}^2+2N\left( {N+1} \right)\hat\phi_{\left\lceil {\frac{{k'}}{2}} \right\rceil}^2+N^2\left(N+1\right)^2\hat\phi_{\left\lceil {\frac{{k'}}{2}} \right\rceil}^2
\end{split}
\end{equation}
\begin{equation}\label{var_down_ext2}
\begin{split}
&{\rm E}\left[ {{\hat{\bf g}}_{k'}^T\left( {{\hat{\bf g}}_{2i - 1}^*{\hat{\bf g}}_{2i}^H + {\hat{\bf g}}_{2i}^*{\hat{\bf g}}_{2i - 1}^H} \right){{\hat{\bf g}}_k}{\hat{\bf g}}_k^H\left( {{{\hat{\bf g}}_{2i - 1}}{\hat{\bf g}}_{2i}^T + {{\hat{\bf g}}_{2i}}{\hat{\bf g}}_{2i - 1}^T} \right){\hat{\bf g}}_{k'}^*} \right]= 2{\rm E}\left[ {{\hat{\bf g}}_{2i - 1}^H{{\hat{\bf g}}_k}{\hat{\bf g}}_k^H{{\hat{\bf g}}_{2i - 1}}} \right]{\rm E}\left[ {{\hat{\bf g}}_{2i}^H{{\hat{\bf g}}_{k'}}{\hat{\bf g}}_{k'}^H{{\hat{\bf g}}_{2i}}} \right]\\
&~~+ 2{\rm Tr}\left\{ {{\rm E}\left[ {{{\hat{\bf g}}_{k'}}{\hat{\bf g}}_{k'}^H} \right]{\rm E}\left[ {{{\hat{\bf g}}_{2i}}{\hat{\bf g}}_{2i}^H} \right]{\rm E}\left[ {{{\hat{\bf g}}_k}{\hat{\bf g}}_k^H} \right]{\rm E}\left[ {{{\hat{\bf g}}_{2i - 1}}{\hat{\bf g}}_{2i - 1}^H} \right]} \right\}=~2N\left(N+1\right)\hat\phi_{i}\hat\phi_{\left\lceil {\frac{{k'}}{2}} \right\rceil}
\end{split}
\end{equation}
\begin{subequations}
    \label{var_down_main2}
    \begin{align}
    \begin{split}
    {{\rm E}\left[ {{\hat{\bf g}}_{k'}^T{\bf F}{{{\bf \xi}}_k}{{\bf \xi}}_k^H{\bf F}^H{\hat{\bf g}}_{k'}^*} \right]}=N\left( {N + 1} \right)^2{\alpha_1 ^2}\sigma _{{{\bf{\xi }}_{k}}}^2\hat \sigma _{k'}^2\hat\phi_{\left\lceil {\frac{{k'}}{2}} \right\rceil} + 2N\left( {N + 1} \right){\alpha_1 ^2}\sigma _{{{\bf{\xi }}_{k}}}^2\hat \sigma _{k'}^2\hat \Phi
    \end{split}\\
    \begin{split}
    {{\rm E}\left[ {{{\bf \xi}}_{k'}^T{\bf F}{{\hat{\bf g}}_k}{\hat{\bf g}}_k^H{\bf F}^H{{\bf \xi}}_{k'}^*} \right]}=N\left( {N + 1} \right)^2{\alpha_1 ^2}\sigma _{{{\bf{\xi }}_{k'}}}^2\hat \sigma _{k}^2\hat\phi_{\left\lceil {\frac{{k'}}{2}} \right\rceil} + 2N\left( {N + 1} \right){\alpha_1 ^2}\sigma _{{{\bf{\xi }}_{k'}}}^2\hat \sigma _{k}^2\hat \Phi
    \end{split}\\
    \begin{split}
    {{\rm E}\left[ {{{\bf \xi}}_{k'}^T{\bf F}{{{\bf \xi}}_k}{{\bf \xi}}_k^H{\bf F}^H{{\bf \xi}}_{k'}^*} \right]}=2N\left( {N + 1} \right){\alpha_1 ^2}\sigma _{{{\bf{\xi }}_k}}^2\sigma _{{{\bf{\xi }}_{k'}}}^2\hat \Phi.
    \end{split}
    \end{align}
\end{subequations}
\begin{equation}\label{var_down}
\begin{split}
{\rm Var}\left[ {{{\bf g}}_{k'}^T{\bf{F}}{{{\bf g}}_k}} \right] = &N\left( {N+1} \right){\alpha_1 ^2}\left[\left( {N + 1} \right)\hat\phi_{\left\lceil {\frac{{k'}}{2}} \right\rceil}\left(\sigma _{k'}^2\hat \sigma _{k}^2+\sigma _k^2\hat \sigma _{k'}^2\right)+2\phi_{\left\lceil {\frac{{k'}}{2}} \right\rceil}\hat\Phi\right]
\end{split}
\end{equation}
\begin{subequations}
    \label{down_others}
    \begin{align}
    \begin{split}
    {\rm{S}}{{\rm{I}}_{k'}}  = & p_{k'}4N\left( {N + 1} \right){\alpha_1 ^2}\sigma _{{{\bf{\xi }}_{k'}}}^2\left[ {\left( {N + 1} \right)\hat \sigma _k^2\hat \sigma _{k'}^4 + \left(\sigma _{k'}^2+\hat\sigma _{k'}^2\right)\hat \Phi } \right]
    \end{split}\\
    \begin{split}
    {\rm{I}}{{\rm{P}}_{k'}}  = &\sum\limits_{i \ne k,k'}^{2K} {{p_i} N\left( {N + 1} \right){\alpha_1 ^2}\left[ {\left( {N + 1} \right)\left( {\sigma _i^2\hat\sigma _{k'}^4\hat\sigma _{k}^2 + \sigma _{k'}^2\hat\sigma _{i}^4\hat\sigma _{i'}^2} \right) + 2\sigma _i^2\sigma _{k'}^2\hat\Phi } \right]}
    \end{split}\\
    \begin{split}
    {\rm{N}}{{\rm{R}}_{k'}}  = &N\left( {N + 1} \right){\alpha_1 ^2}\sigma _n^2\left[ {\left( {N + 1} \right)\hat\sigma _k^2\hat\sigma _{k'}^4 + 2\sigma _{k'}^2\hat\Phi } \right],~{\rm{N}}{{\rm{U}}_{k'}}  = \sigma _n^2.
    \end{split}
    \end{align}
\end{subequations}
\begin{equation}\label{SCSI_rate_alpha}
\begin{split}
&\frac{{{p_k}{{\left| {{\rm E}\left[ {{{\bf g}}_{k'}^T{\bf{F}}{{{\bf g}}_k}} \right]} \right|}^2}}}{{{p_k}{\rm Var}\left[ {{{\bf g}}_{k'}^T{\bf{F}}{{{\bf g}}_k}} \right] + {\rm{S}}{{\rm{I}}_{k'}} + {\rm{I}}{{\rm{P}}_{k'}} + {\rm{N}}{{\rm{R}}_{k'}} + {\rm{N}}{{\rm{U}}_{k'}}}}\\
\mathop  = \limits^{(a)}~&\frac{{{p_k}N\left( {N + 1} \right)\hat\sigma _k^4\hat\sigma _{k'}^4}}{\sum\limits_{i=1}^{2K}{{p_i}\left\{\varsigma_{k',i}+\frac{\sigma_n^2}{P_{\rm R}}\left[2\hat\Phi\sigma_i^2\sigma_{i'}^2+\left(N+1\right)\hat\sigma_i^4\hat\sigma_{i'}^2\right]\right\}}+{c_{k'}}{p_{k'}}+\sigma _n^2\left[ {\left( {N + 1} \right)\hat\sigma _k^2\hat\sigma _{k'}^4 + 2\sigma _{k'}^2\hat\Phi } \right]+ \frac{{2\sigma _n^4\hat\Phi}}{{{P_{\rm{R}}}}}}
\end{split}
\end{equation}
\begin{equation}\label{hat_mean_upp}
\begin{split}
&{\rm E}\left[ {{\bf{g}}_{k'}^T{\bf{F}}{{\bf{g}}_k}} \right] =\alpha_2 {\rm E}\left[ {{{\left( {{{{\hat{\bf g}}}_{k'}} - {{\bf{\xi }}_{k'}}} \right)}^T}{{\hat{ \bar{\bf G}}}^*}{\bf{{ T}}}{{\hat{ \bar{ \bf G}}}^H}\left( {{{{\hat{\bf g}}}_k} - {{\bf{\xi }}_k}} \right)} \right]\mathop  = \limits^{(a)}\alpha_2 {\rm E}\left[ {{\hat{\bf g}}_{k'}^T{{\hat{ \bar{ \bf G}}}^*}{\bf{{T}}}{{\hat{ \bar{ \bf G}}}^H}{{\hat{{\bf g}}}_k}} \right] + \alpha_2 {\rm E}\left[ {{\bf{\xi }}_{k'}^T{{\hat{ \bar{ \bf G}}}^*}{\bf{{T}}}{{\hat{ \bar{ \bf G}}}^H}{{\bf{\xi }}_k}} \right]\mathop  = \limits^{(b)} \alpha_2
\end{split}
\end{equation}
\end{table*}

Then, the variance of ${{{\bf g}}_{k'}^T{\bf{F}}{{{\bf g}}_k}}$ in the denominator of (\ref{SCSI_rate}) is (\ref{var_down_main}), where step $(a)$ indicates the definition of the variance and $(b)$ results from ${{\bf g}}_{i}={\hat{\bf g}}_{i}-{\bf \xi}_{i}$ and the independence between ${\hat{\bf g}}_{i}$ and ${\bf \xi}_{i}$~($\forall i\in\{1,\cdots,2K\}$). For the first term of step $(b)$ in (\ref{var_down_main}), we have (\ref{var_down_main1}), where step $(b)$ is obtained simply by the property ${\rm Tr}\left\{ {{\bf{AB}}} \right\} = {\rm Tr}\left\{ {{\bf{BA}}} \right\}$, the properties of Gaussian distributed vectors, and \cite[Lemma~2.9]{Tulino2004}. To elaborate in detail, for the items of $(a)$ in (\ref{var_down_main1}) with $i = \left\lceil {\frac{{k'}}{2}} \right\rceil$, we have (\ref{var_down_ext1}), where the properties of ${\rm E}\left[ {{{\hat{\bf g}}_{i}}{\hat{\bf g}}_{i}^H{{\hat{\bf g}}_{i}}{\hat{\bf g}}_{i}^H} \right] = {\left( {N + 1} \right)}\sigma _i^4{{\bf{I}}_N}$~($i=1,2,\cdots,2K$) resulting from the fact that vectors ${\hat{\bf g}}_i$ contains the i.i.d. $\mathcal{CN}(0,\sigma_i^2)$ elements, ${\rm E}\left[ {{{\left| \theta  \right|}^4}} \right] = 2\sigma _\theta ^4$ for arbitrary complex value $\theta \sim \mathcal{CN}(0,\sigma_{\theta}^2)$, and \cite[Lemma~2.9]{Tulino2004} are utilized, respectively. While for the items with $i \ne \left\lceil {\frac{{k'}}{2}} \right\rceil$, we have (\ref{var_down_ext2}), where ${\rm Tr}\left\{ {{\bf{AB}}} \right\} = {\rm Tr}\left\{ {{\bf{BA}}} \right\}$ and the properties of Gaussian distributed vectors are utilized, respectively. Subsequently, the left three terms of step $(b)$ in (\ref{var_down_main}) are calculated and expressed as (\ref{var_down_main2}). Substituting (\ref{var_down_main1}) and (\ref{var_down_main2}) into (\ref{var_down_main}) leads to (\ref{var_down}),
where we define $\Phi  = \sum\limits_{i = 1}^K {{\phi _i}}$ with ${\phi _i} = \sigma _{2i - 1}^2 \sigma _{2i}^2$.

Similarly, we obtain (\ref{down_others}).
Substituting (\ref{mean_upp}), (\ref{var_down}) and (\ref{down_others}) into (\ref{SCSI_rate}), we have (\ref{SCSI_rate_alpha}), where $\varsigma_{k',i}\buildrel \Delta \over =  {\left( {N + 1} \right)\left({\sigma _i^2\hat\sigma _{k'}^4\hat\sigma _{k}^2 + \sigma _{k'}^2\hat\sigma _{i}^4\hat\sigma _{i'}^2}\right) + 2\sigma _i^2\sigma _{k'}^2\hat\Phi }$, $c_{k'}=2\left[ \left( {N + 1} \right)\left(\sigma_{k'}^2-2\hat\sigma_{k'}^2\right)\hat\sigma _k^2\hat\sigma _{k'}^4 + \left(\sigma_{k'}^4\right.\right.$ $\left.\left. -2\hat\sigma_{k'}^4\right)\hat\Phi \right]$, and step $(a)$ is obtained by substituting (\ref{alpha}) into (\ref{SCSI_rate_alpha}). In this way, (\ref{SCSI_rate_CF}) is obtained and thus Theorem 1 is demonstrated.

\section*{Appendix D: Proof of Theorem 2}\label{Appendix_D}
In this appendix, we focus on the proof of the closed-form expression in (\ref{SCSI_rate}) with imperfect CSI based ZFR/ZFT processing. First, we start from the expectation ${\rm E}\left[ {{\bf{g}}_{k'}^T{\bf{F}}{{\bf{g}}_k}} \right]$ in the numerator, given by~(\ref{hat_mean_upp}),
where step $(a)$ is based on the fact that the estimation error matrix ${\bf{\Xi}}$ is independent of ${{\bf{\hat G}}}$, i.e., ${\bf{\xi}}_i$ is independent of ${{\bf{\hat g}}}_j$ for $\forall i,j\in \left\{1,2,\cdots,2K\right\}$, and step $(b)$ is obtained by ${\hat{\bf{g}}_{i'}^T{\bf{F}}{\hat{\bf{g}}_j}}=\alpha_2{\delta _{ij}}$ on account of ${\hat{\bf{G}}^T}{\bf{F}}\hat{\bf{G}} = \alpha_2 {\bf I}_{2K}$.

Then, based on the imperfect CSI based ZFR/ZFT processing in (\ref{hat_alpha}), the variance of ${{\bf{g}}_{k'}^T{\bf{F}}{{\bf{g}}_k}}$ in the denominator of (\ref{SCSI_rate}) is (\ref{hat_var_down}),
\begin{table*}
\begin{equation}\label{hat_var_down}
\begin{split}
&{\rm Var}\left[ {{\bf{g}}_{k'}^T{\bf{F}}{{\bf{g}}_k}} \right] = {\rm E}\left[ {{{\left| {{\bf{g}}_{k'}^T{\bf{F}}{{\bf{g}}_k}} \right|}^2}} \right] - {\left| {{\rm E}\left[ {{\bf{g}}_{k'}^T{\bf{F}}{{\bf{g}}_k}} \right]} \right|^2} = {\rm E}\left[ {{\bf{\hat g}}_{k'}^T{\bf{F}}{{{\bf{\hat g}}}_k}{\bf{\hat g}}_k^H{{\bf{F}}^H}{\bf{\hat g}}_{k'}^*} \right]- {{\alpha_2 }^2} + {\rm E}\left[ {{\bf{\hat g}}_{k'}^T{\bf{F}}{{\bf{\xi }}_k}{\bf{\xi }}_k^H{{\bf{F}}^H}{\bf{\hat g}}_{k'}^*} \right]\\
&+ {\rm E}\left[ {{\bf{\xi }}_{k'}^T{\bf{F}}{{{\bf{\hat g}}}_k}{\bf{\hat g}}_k^H{{\bf{F}}^H}{\bf{\xi }}_{k'}^*} \right] + {\rm E}\left[ {{\bf{\xi }}_{k'}^T{\bf{F}}{{\bf{\xi }}_k}{\bf{\xi }}_k^H{{\bf{F}}^H}{\bf{\xi }}_{k'}^*} \right] \mathop  = \limits^{(a)}\sigma_{{\xi _k}}^2{\rm E}\left[ {{\bf{\hat g}}_{k'}^T{\bf{F}}{{\bf{F}}^H}{\bf{\hat g}}_{k'}^*} \right] + \sigma _{{\xi _{k'}}}^2{\rm E}\left[ {{\bf{\hat g}}_k^H{{\bf{F}}^H}{\bf{F}}{{{\bf{\hat g}}}_k}} \right]\\
&+ \sigma _{{\xi _k}}^2\sigma _{{\xi _{k'}}}^2{\rm Tr}\left\{ {{\rm E}\left[ {{\bf{F}}{{\bf{F}}^H}} \right]} \right\} \mathop = \limits^{(b)}{{\alpha_2 }^2}\sigma _{{\xi _k}}^2{\rm E}\left[ {{\bf{e}}_{k'}^T{\bf T}\hat{\bf{\Omega}}{\bf T}{{\bf{e}}_{k'}}} \right] + {{\alpha_2 }^2}\sigma _{{\xi _{k'}}}^2{\rm E}\left[ {{\bf{e}}_k^T{\bf T}\hat{\bf{\Omega}}^*{\bf T}{{\bf{e}}_k}} \right] + {{\alpha_2 }^2}\sigma _{{\xi _k}}^2\sigma _{{\xi _{k'}}}^2{\rm Tr}\left\{ {{\rm E}\left[ {{{\hat{\bf{ \Omega}}}^*}{\bf T}\hat{\bf{\Omega}}{\bf T}} \right]} \right\}\\
&\mathop  = \limits^{(c)}{{\alpha_2 }^2}\sigma _{{\xi _k}}^2{\rm E}\left[ {{{\hat \omega}_{k,k}}} \right] + {{\alpha_2 }^2}\sigma _{{\xi _{k'}}}^2{\rm E}\left[ {\hat \omega_{k',k'}^*} \right] + {{\alpha_2 }^2}\sigma _{{\xi _k}}^2\sigma _{{\xi _{k'}}}^2\sum\limits_{j = 1}^{2K} {\left( {{\rm E}\left[ {\hat \omega_{j,j}^*{{\hat \omega}_{j',j'}}} \right] + {\rm E}\left[ {{{\left| {{{\hat \omega }_{j,j'}}} \right|}^2}} \right]} \right)}\mathop  = \limits^{(d)}{{\alpha_2 }^2}{\theta _{k',k}}
\end{split}
\end{equation}
\end{table*}
where ${\theta _{i,j}}  = \frac{{\sigma _{{\xi _j}}^2}}{{\left( {N - 2K - 1} \right)\hat \sigma _{i'}^2}} + \frac{{\sigma _{{\xi _i}}^2}}{{\left( {N - 2K - 1} \right)\hat \sigma _{j'}^2}} + \sigma _{{\xi _i}}^2\sigma _{{\xi _j}}^2\hat \eta$ with $i,j \in \left\{ {1,2, \cdots, 2K} \right\}$, step $(a)$ results from the property ${\rm Tr}\left\{ {{\bf{AB}}} \right\} = {\rm Tr}\left\{ {{\bf{BA}}} \right\}$, ${{\bf{\hat g}}_{k'}^T{\bf{F}}{{{\bf{\hat g}}}_k}{\bf{\hat g}}_k^H{{\bf{F}}^H}{\bf{\hat g}}_{k'}^*}={\alpha_2}^2\delta_{k,k}\delta_{k',k'}$ and ${\rm E}\left[ {\bf{\xi }}_{i}{\bf{\xi }}_{i}^H \right]=\sigma_{\xi_i}^2{\bf I}_{N}$ for $\forall i\in\left\{1,2,\cdots,2K\right\}$, step $(b)$ is obtained by the definition of ${\hat {\bf \Omega }} \buildrel \Delta \over = {{\left( {{\hat {\bf{G}}^H}\hat {\bf{G}}} \right)}^{ - 1}}$ with $\hat {\omega }_{i,j} = {\left( \hat {\bf{\Omega }} \right)_{i,j}}$ for $\forall i,j\in\left\{1,2,\cdots,2K\right\}$ and the fact of ${{\bf{g}}_{k'}^T{{\bf{G}}^*}{{\left( {{{\bf{G}}^T}{{\bf{G}}^*}} \right)}^{ - 1}}}={\bf{e}}_{k'}^T$ and ${{{\left( {{{\bf{G}}^T}{{\bf{G}}^*}} \right)}^{ - 1}}{{\bf{G}}^T}{\bf{g}}_{k'}^*}={{\bf{e}}_{k'}}$, $(c)$ is just an intuitive transformation, and $(d)$ is derived according to the properties of the inverse Wishart matrix in (\ref{hat_alpha2_ext1}) and (\ref{hat_alpha2_ext2}).

Considering that no SIC is performed as ${\hat{\bf g}}_{k'}^T{\bf{F}}{{ {\hat{\bf g}}}_{k'}}=0$, the self-interference term can be rewritten as
\begin{equation}\label{hat_down_others1}
    \begin{split}
    &{\rm{S}}{{\rm{I}}_{k'}} =p_{k'}{\rm E}\left[ {{{\left| {{\bf{g}}_{k'}^T{\bf{F}}{{\bf{g}}_{k'}}}\right|}^2}} \right]=p_{k'}{\alpha_2 ^2}\left({\theta _{k',k'}} + \sigma _{{\xi _{k'}}}^4\hat \eta\right).
    \end{split}
\end{equation}

Similarly, we obtain
\begin{subequations}
    \label{hat_down_others2}
    \begin{align}
    \begin{split}
    {\rm{I}}{{\rm{P}}_{k'}} &  =  \sum\limits_{i \ne k,k'}^{2K} {{p_i}{{\alpha_2 }^2}{\theta _{k',i}}},~{\rm{N}}{{\rm{U}}_{k'}} = \sigma _n^2
    \end{split}\\
    \begin{split}
    {\rm{N}}{{\rm{R}}_{k'}} &  = \frac{{{{\alpha_2 }^2}\sigma _n^2}}{{\left( {N - 2K - 1} \right)\hat \sigma _k^2}} + {{\alpha_2 }^2}\sigma _n^2\sigma _{{\xi _{k'}}}^2\hat \eta.
    \end{split}
    \end{align}
\end{subequations}
Substituting (\ref{hat_mean_upp}), (\ref{hat_var_down}), (\ref{hat_down_others1}) and (\ref{hat_down_others2}) into (\ref{SCSI_rate}), we have (\ref{hat_SCSI_rate_alpha2}),
\begin{table*}
\begin{equation}\label{hat_SCSI_rate_alpha2}
\begin{split}
&\frac{{{p_k}{{\left| {{\rm E}\left[ {{\bf{g}}_{k'}^T{\bf{F}}{{\bf{g}}_k}} \right]} \right|}^2}}}{{{p_k}{\rm Var}\left[ {{\bf{g}}_{k'}^T{\bf{F}}{{\bf{g}}_k}} \right] + {\rm{S}}{{\rm{I}}_{k'}}+ {\rm{I}}{{\rm{P}}_{k'}} + {\rm{N}}{{\rm{R}}_{k'}} + {\rm{N}}{{\rm{U}}_{k'}}}}\\
\mathop  = \limits^{(a)}~&\frac{{{p_k}}}{{\sum\limits_{i = 1}^{2K} {{p_i}\left[ {{\theta _{k',i}} + \frac{{\sigma _n^2}}{{{P_{\rm{R}}}}}\left( {\frac{1}{{\left( {N - 2K - 1} \right)\hat \sigma _{i'}^2}} + \sigma _{{\xi _i}}^2\hat \eta } \right)} \right]}  + {p_{k'}}\sigma _{{\xi _{k'}}}^4\hat \eta + \frac{{\sigma _n^2}}{{\left( {N - 2K - 1} \right)\hat \sigma _k^2}} + \sigma _n^2\sigma _{{\xi _{k'}}}^2\hat \eta  + \frac{{\sigma _n^4\hat \eta }}{{{P_{\rm{R}}}}}}}
\end{split}
\end{equation}
\end{table*}
where $(a)$ results form substituting (\ref{hat_alpha}) into (\ref{hat_SCSI_rate_alpha2}). Thus (\ref{hat_SCSI_rate_CF}) is obtained and Theorem 2 is demonstrated.

\end{document}